\documentclass[letterpaper]{article} 
\def\year{2022}\relax
\usepackage{aaai22}  
\usepackage{times}  
\usepackage{helvet}  
\usepackage{courier}  
\usepackage[hyphens]{url}  
\usepackage{graphicx} 
\usepackage{hyperref}
\usepackage{natbib}  
\usepackage{caption} 
\DeclareCaptionStyle{ruled}{labelfont=normalfont,labelsep=colon,strut=off} 
\usepackage[absolute,overlay]{textpos} 

\usepackage{multirow}
\usepackage{newfloat}
\usepackage{listings}
\newcommand{\liu}[1]{\textcolor{black}{#1}}

\usepackage{enumerate}
\usepackage{cite}
\usepackage{amsmath,amssymb,amsfonts}
\usepackage{mathrsfs}
\usepackage{textcomp}
\usepackage{xcolor}
\usepackage{graphicx}
\usepackage{algorithm}
\usepackage{algpseudocode}
\usepackage{booktabs}
\usepackage{cleveref}
\usepackage[labelformat=simple]{subcaption}

\newcommand{\argmin}{\mathop{\mathrm{argmin}}}

\title{Efficient Device Scheduling with Multi-Job Federated Learning}
\author{Chendi Zhou$^{1}$\setcounter{footnote}{1}\footnote{C. Zhou and J. Liu contributed equally to the paper. This work was done when C. Zhou was an intern at Baidu Inc.}, Ji Liu$^{2}$\footnotemark[2]\setcounter{footnote}{0}\footnote{Corresponding author (liuji04@baidu.com).}, Juncheng Jia$^1$, Jingbo Zhou$^2$, Yang Zhou$^3$, Huaiyu Dai$^4$, Dejing Dou$^2$}
\affiliations{
\textsuperscript{\rm 1}Soochow University, 
\textsuperscript{\rm 2}Baidu Inc., China \\
\textsuperscript{\rm 3}Auburn University,
\textsuperscript{\rm 4}North Carolina State University, United States
}

\usepackage{bibentry}

\begin{document}

\begin{textblock*}{10cm}(7cm,26cm) 
   {\huge To appear in AAAI 2022}
\end{textblock*}

\maketitle

\begin{abstract}
Recent years have witnessed a large amount of decentralized data in multiple (edge) devices of end-users, while the aggregation of the decentralized data remains difficult for machine learning jobs due to laws or regulations. 
Federated Learning (FL) emerges as an effective approach to handling decentralized data without sharing the sensitive raw data, while collaboratively training global machine learning models. 
The servers in FL need to select (and schedule) devices during the training process. 
However, the scheduling of devices for multiple jobs with FL remains a critical and open problem.
In this paper, we propose a novel multi-job FL framework to enable the parallel training process of multiple jobs. 
The framework consists of a system model and two scheduling methods.
In the system model, we propose a parallel training process of multiple jobs, and construct a cost model based on the training time and the data fairness of various devices during the training process of diverse jobs. 
We propose a reinforcement learning-based method and a Bayesian optimization-based method to schedule devices for multiple jobs while minimizing the cost.
We conduct extensive experimentation with multiple jobs and datasets. 
The experimental results show that our proposed approaches significantly outperform baseline approaches in terms of training time (up to 8.67 times faster) and accuracy (up to 44.6\% higher). 
\end{abstract}


\section{Introduction}

Recent years have witnessed a large amount of decentralized data over various Internet of Things (IoT) devices, mobile devices, etc. \cite{liu2021distributed}, which can be exploited to train machine learning models of high accuracy for diverse artificial intelligence applications. Since the data contain sensitive information of end-users, a few stringent legal restrictions \cite{GDPR, CCL, CCPA, chik2013singapore} have been put into practice to protect data security and privacy. In this case, it is difficult or even impossible to aggregate the decentralized data into a single server or a data center to train machine learning models. To enable collaborative training with distributed data, federated learning (FL) \cite{mcmahan2017communication}, which does not transfer raw data, emerges as an effective approach.

FL was first introduced to collaboratively train a global model with non-Independent and Identically Distributed (non-IID) data distributed on mobile devices \cite{mcmahan2017communication}. 
During the training process of FL, the raw data is kept decentralized without being moved to a single server or a single data center \cite{kairouz2019advances, yang2019federated}.
FL only allows the intermediate data to be transferred from the distributed devices, which can be the weights or the gradients of a model. 
FL generally exploits a parameter server architecture \cite{smola2010architecture}, where a server (or a group of servers) coordinates the training process with numerous devices. 
To collaboratively train a global model, the server selects (schedules) a number of devices to perform local model updates based on their local data, and then it aggregates the local models to obtain a new global model. 
This process is repeated multiple times so as to generate a global model of high accuracy.

While current FL solutions \cite{mcmahan2017communication, pilla2021optimal} focus on a single-task job or a multi-task job \cite{Smith2017Multi-Task}, FL with multiple jobs \cite{han2020marble} remains an open problem. The major difference between the multi-task job and multiple jobs is that the tasks of the multi-task job share some common parts of the model, while the multiple jobs do not have interaction between each other in terms of the model. The multi-job FL deals with the simultaneous training process of multiple independent jobs. Each job corresponds to multiple updates during the training process of a global model with the corresponding decentralized data. While the FL with a single job generally chooses a portion of devices to update the model, the other devices remain idle, and the efficiency is low. The multi-job FL can well exploit diverse devices for multiple jobs simultaneously, which brings high efficiency. The available devices are generally heterogeneous \cite{li2020federated, Li2021Heterogenous}, i.e., the computing and communication capability of each device is different, and the data in each device may also differ. During the training process of multiple jobs, the devices need to be scheduled to each job. At a given time, a device can be scheduled to only one job. However, only a portion of the available devices is scheduled to one job in order to reduce the influence of stragglers \cite{mcmahan2017communication}. Powerful devices should be scheduled to jobs in order to accelerate the training process, while other eligible devices should also participate in the training process to increase the fairness of data so as to improve the accuracy of the final global models. The fairness of data refers to the fair participation of the data in the training process of FL, which can be indicated by the standard deviation of the times to be scheduled to a job \cite{pitoura2007load,finkelstein2008fairness}.

While the scheduling problem of devices is typical NP-hard \cite{Du1989NPHard, liu2020job}, some solutions have already been proposed for the training process of FL \cite{McMahan2017Communication-efficien, Nishio2019Client, Li2021Heterogenous, Abdulrahman2021FedMCCS} or distributed systems \cite{barika2019scheduling}, which generally only focus on a single job with FL. In addition, these methods either cannot address the heterogeneity of devices \cite{McMahan2017Communication-efficien}, or do not consider the data fairness during the training process \cite{Nishio2019Client, Li2021Heterogenous, Abdulrahman2021FedMCCS}, which may lead to low accuracy. 

In this paper, we propose a Multi-Job Federated Learning (MJ-FL) framework to enable the efficient training of multiple jobs with heterogeneous edge devices. The MJ-FL framework consists of a system model and two scheduling methods. The system model enables the parallel training process of multiple jobs. With the consideration of both the efficiency of the training process, i.e., the time to execute an iteration, and the data fairness of each job for the accuracy of final models, we propose a cost model based on the training time and the data fairness within the system model. We propose two scheduling methods, i.e., reinforcement learning-based and Bayesian optimization-based, to schedule the devices for each job. To the best of our knowledge, we are among the first to study FL with multiple jobs. We summarize our contributions as follows:

\begin{itemize}
    \item We propose MJ-FL, a multi-job FL framework consisting of a parallel training process for multiple jobs and a cost model for the scheduling methods.
    We propose combining the capability and data fairness in the cost model to improve the efficiency of the training process and the accuracy of the global model.
    \item We propose two scheduling methods, i.e., Reinforcement Learning (RL)-based and Bayesian Optimization (BO)-based methods, to schedule the devices to diverse jobs. Each method has advantages in a specific situation. The BO-based method performs better for simple jobs, while the RL-based method is more suitable for complex jobs.
    \item We carry out extensive experimentation to validate the proposed approach. We exploit multiple jobs, composed of Resnet18, CNN, AlexNet, VGG, and LeNet, to demonstrate the advantages of our proposed approach using both IID and non-IID datasets.
\end{itemize}

The rest of the paper is organized as follows. We present the related work in Section \ref{sec:related_work}. Then, we explain the system model and formulate the problem with a cost model in Section \ref{sec:system_model}. We present the scheduling methods in Section \ref{sec:solutions}. The experimental results with diverse models and datasets are given in Section \ref{sec:experiment}. Finally, Section \ref{sec:conclusion} concludes the paper.

\section{Related Work}
\label{sec:related_work}

In order to protect the security and privacy of decentralized raw data, FL emerges as a promising  approach, which enables training a global model with decentralized data  \cite{mcmahan2017communication, yang2019federated, li2020federated, liu2021distributed}. Based on the data distribution, FL can be classified into three types, i.e., horizontal, vertical, and hybrid \cite{yang2019federated, liu2021distributed}. The horizontal FL addresses the decentralized data of the same features, while the identifications are different. The vertical FL handles the decentralized data of the same identifications with different features. The hybrid FL deals with the data of different identifications and different features. In addition, FL includes two variants: cross-device FL and cross-silo FL \cite{kairouz2019advances}. The cross-device FL trains global machine learning models with a huge number of mobile or IoT devices, while the cross-silo FL handles the collaborative training process with the decentralized data from multiple organizations or geo-distributed datacenters. In this paper, we focus on the horizontal and cross-device FL.

Current FL approaches \cite{Bonawitz19, liu2020fedvision, yurochkin2019bayesian,  Wang2020Federated} generally deal with a single job, i.e., with a single global model. While some FL approaches have been proposed to handle multiple tasks \cite{Smith2017Multi-Task, chen2021matching}, the tasks share some common parts of a global model and deal with the same types of data. In addition, the devices are randomly selected (scheduled) in these approaches. 

A few scheduling approaches \cite{McMahan2017Communication-efficien, Nishio2019Client, Li2021Heterogenous, Abdulrahman2021FedMCCS, barika2019scheduling, Nishio2019Client, Li2021Heterogenous, Abdulrahman2021FedMCCS, sun2020deepweave} exist for single-job scheduling while the device scheduling with multi-job FL is rarely addressed. The scheduling methods in the above works are mainly based on some heuristics. For instance, the greedy method \cite{shi2020device} and the random scheduling method \cite{McMahan2017Communication-efficien} are proposed for FL, while genetic algorithms \cite{barika2019scheduling} are exploited for distributed systems. However, these methods do not consider the fairness of data, which may lead to low accuracy for multi-job FL. \liu{The black-box optimization-based methods, e.g., RL \cite{sun2020deepweave}, BO \cite{kim2020probabilistic}, and deep neural network \cite{zang2019hybrid}, have been proposed to improve the efficiency, i.e., the reduction of execution time, in distributed systems.} They do not consider data fairness either, which may lead to low accuracy for multi-job FL.

Different from all existing works, we propose a system model for the multi-job FL with the consideration of both efficiency and accuracy. In addition, we propose two scheduling methods, one based on RL and the other based on BO, for multi-job FL, which are suitable for diverse models and for both IID and non-IID datasets.

\section{System Model and Problem Formulation}
\label{sec:system_model}

In this section, we first explain the motivation for multi-job FL. Then, we propose our multi-job FL framework 
, consisting of multi-job FL process and a cost model. Afterward, we formally define the problem to address in this paper.

\begin{figure*}[htbp]
\centerline{\includegraphics[width=0.8\linewidth]{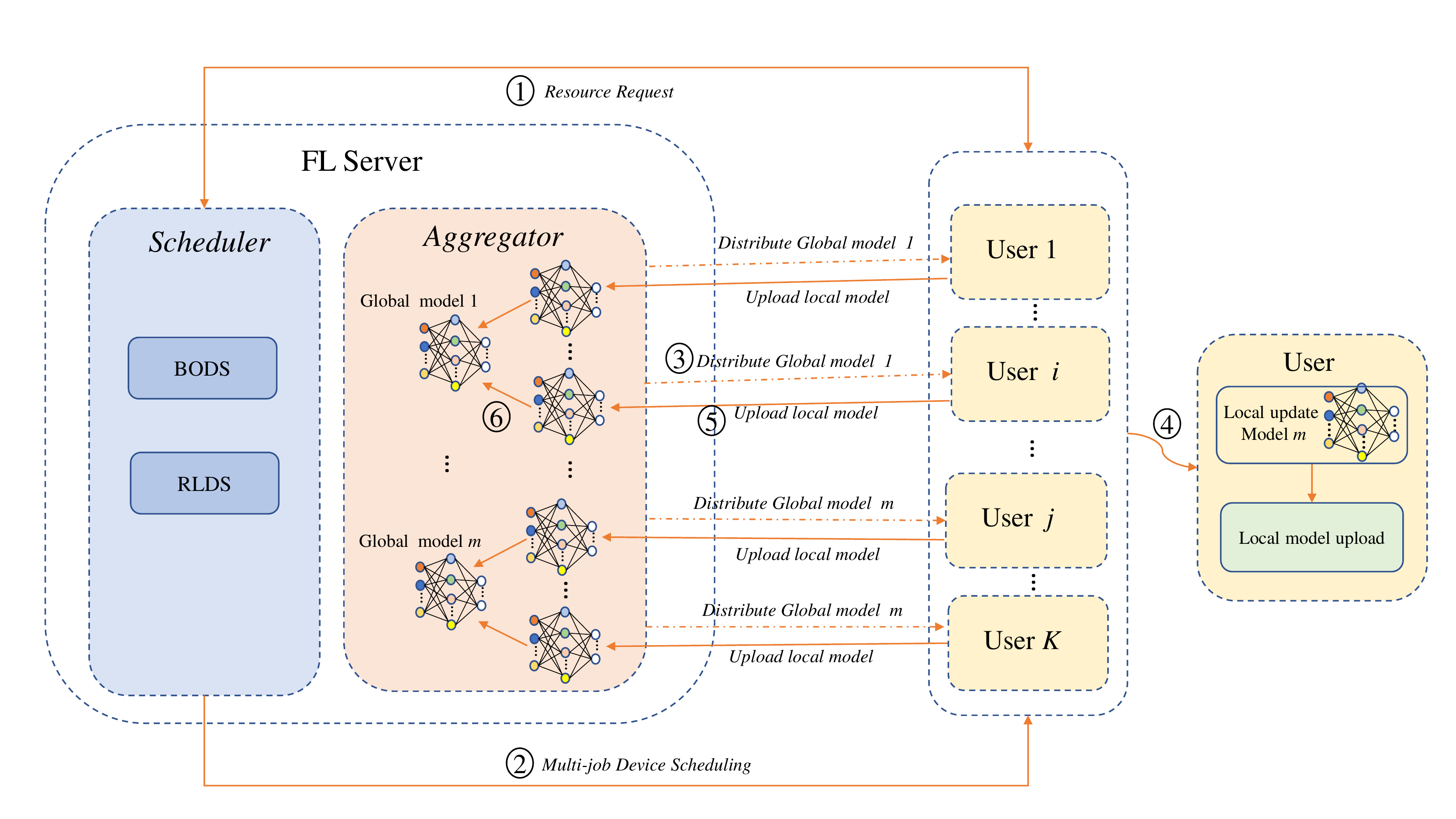}}
\vspace{-4mm}
\caption{The training process within the Multi-job Federated Learning Framework.}
\label{framework}
\vspace{-3mm}
\end{figure*}

\subsection{Motivation for Multi-Job Federated Learning}
Let us assume a scenario where there are multiple FL jobs to be processed at the same time, e.g., image classification, speech recognition, and text generation. These jobs can be trained in parallel so as to efficiently exploit the available devices. However, while each device can only update the model of one job at a given time slot, it is critical to schedule devices to different jobs during the training process. 
As the devices are generally heterogeneous, some devices may possess high computation or communication capability while others may not. 
In addition, the data fairness of multiple devices may also impact the convergence speed of the training process. 
For instance, if only certain powerful devices are scheduled to a job, the model can only learn the knowledge from the data stored on these devices, while the knowledge from the data stored on other devices may be missed. In order to accelerate the training process of multiple jobs with high accuracy, it is critical to consider how to schedule devices while taking into consideration both the computing and communication capability and the data fairness.

A straightforward approach is to train each job separately using the mechanism explained in \cite{McMahan2017Communication-efficien}, while exploiting the existing scheduling of single-job FL, e.g., FedAvg \cite{McMahan2017Communication-efficien}.
In this way, simple parallelism is considered while the devices are not fully utilized and the system is of low efficiency. 
In addition, a direct adaptation of existing scheduling methods to multi-job FL cannot address the efficiency and the accuracy at the same time.
Thus, it is critical to propose a reasonable and effective approach for the multi-job FL.

\subsection{Multi-job Federated Learning Framework}

In this paper, we focus on an FL environment composed of a server module and multiple devices. The server module (Server) may consist of a single parameter server or a group of parameter servers \cite{li2014scaling}. In this section, we present a multi-job FL framework, which is composed of a process for the multi-job execution and a cost model to estimate the cost of the execution.

\subsubsection{Multi-job FL Process}

Within the multi-job FL process, we assume that $K$ devices, denoted by the set $\mathcal{K}$, collaboratively train machine learning models for $M$ jobs, denoted by the set $\mathcal{M}$. Each device $k$ is assumed to have $M$ local datasets corresponding to the $M$ jobs without loss of generality, and the dataset of the $m$-th job on device $k$ is expressed as $\mathcal{D}_k^m = \{\boldsymbol{x}_{k,d}^m \in \mathbb{R}^{n_m}, y_{k,d}^m \in \mathbb{R} \}_{d=1}^{D_k^m}$ with $D_k^m = |\mathcal{D}_k^m|$ as the number of data samples, $\boldsymbol{x}_{k,d}^m$ representing the $d$-th $n_m$-dimentional input data vector of Job $m$ at Device $k$, and $y_{k,d}^m$ denoting the labeled output of $\boldsymbol{x}_{k,d}^m$. The whole dataset of Job $m$ is denoted by $\mathcal{D}^m = \bigcup_{k \in \mathcal{K}} \mathcal{D}_k^m$ with $D^m=\sum_{k \in \mathcal{K}} D_k^m$.
The objective of multi-job FL is to learn respective model parameters $\{\boldsymbol{w}^m\}$ based on the decentralized datasets. 
The global learning problem of multi-job FL can be expressed by the following formulation:
\vspace{-2mm}
\begin{equation}\label{eq:eq1}
\min_{\boldsymbol{W}}\sum_{m=1}^M\mathbb{L}_m\textrm{, with }\mathbb{L}_m = \sum_{k=1}^K\frac{D_k^m}{D^m}F_k^m(\boldsymbol{w}^m),
\vspace{-2mm}
\end{equation}
where $\mathbb{L}_m$ is the loss value of Job $m$, $F_k^m(\boldsymbol{w}^m)=\frac{1}{D_k^m}\sum_{\{\boldsymbol{x}_{k,d}^m,y_{k,d}^m\}\in\mathcal{D}_k^m} f^m(\boldsymbol{w}^m;\boldsymbol{x}_{k,d}^m,y_{k,d}^m)$ is the loss value of Job $m$ at Device $k$, $\boldsymbol{W}: \equiv\{\boldsymbol{w}^1,\boldsymbol{w}^2,...,\boldsymbol{w}^M\}$ is the set of weight vectors for all jobs, and $f^m(\boldsymbol{w}^m;\boldsymbol{x}_{k,d}^m, y_{k,d}^m)$ captures the error of the model parameter $\boldsymbol{w}^m$ on the data pair $\{\boldsymbol{x}_{k,d}^m, y_{k,d}^m\}$. 

In order to solve the problem defined in Formula \ref{eq:eq1}, the Server needs to continuously schedule  devices for different jobs to update the global models iteratively until the training processes of the corresponding job converge or achieve a target performance requirement (in terms of accuracy or loss value). We design a multi-job FL process as shown in Fig. \ref{framework}. The Server first initializes a global model for each job. The initialization can be realized randomly or from the pre-training process with public data. In order to know the current status of devices, the Server sends requests to available devices in Step \textcircled{1}. Then, in Step \textcircled{2}, the Server schedules devices to the current job, according to a scheduling plan generated from a scheduling method (see details in Section \ref{sec:solutions}). The scheduling plan is a set of devices that are selected to perform the local training process for the current job. Please note that the scheduling process generates a scheduling plan for each job during the training process of multiple jobs, i.e., with an online strategy, while the scheduling processes of multiple jobs are carried out in parallel. The Server distributes the latest global model of the current job to the scheduled devices in Step \textcircled{3}, and then the model is updated in each device based on the local data in Step \textcircled{4}. Afterward, each device uploads the updated model to the Server after its local training in Step \textcircled{5}. Finally, Server aggregates the models of scheduled devices to generate a new global model in Step \textcircled{6}. The combination of Steps \textcircled{1} - \liu{\textcircled{6}} is denoted by a round, which is repeated for each job until the corresponding global model reaches the expected performance (accuracy, loss value, or convergence). Please note that multiple jobs are executed in parallel asynchronously, while a device can only be scheduled to one job at a given time. In addition, we assume that the importance of each job is the same.  

\subsubsection{Cost Model}

In order to measure the performance of each round, we exploit a cost model defined in Formula \ref{eq:totalCost}, which is composed of time cost and data fairness cost. The data fairness has a significant impact on convergence speed.
\vspace{-1mm}
\begin{equation}\label{eq:totalCost}
Cost_{m}^{r}(\mathcal{V}_m^r) = \alpha * \mathscr{T}_m^{r}(\mathcal{V}_m^r) + \beta * \mathscr{F}_m^{r}(\mathcal{V}_m^r),
\vspace{-1mm}
\end{equation}
where $\alpha$ and $\beta$ are the weights of time cost and fairness cost respectively, $\mathscr{T}_m^{r}(\cdot)$ represents the execution time of the training process in Round $r$ with the set of scheduled devices $\mathcal{V}_m^r$, and $\mathscr{F}_m^{r}(\cdot)$ is the corresponding data fairness cost. 

As defined in Formula \ref{eq:time}, the execution time of a round depends on the slowest device in the set of scheduled devices. 
\vspace{-3mm}
\begin{equation}\label{eq:time}
\mathscr{T}_{m}^{r}(\mathcal{V}_m^r) = \max_{k \in \mathcal{V}_m^r}\{t_m^k\},
\vspace{-1mm}
\end{equation}
where $t_m^k$ is the execution time of Round $r$ in Device $k$ for Job $m$. $t_m^k$ is composed of the communication time and the computation time, which is complicated to estimate and differs for different devices. In this study, we assume that the execution time of each device follows the shift exponential distribution as defined in Formula \ref{eq:distribution} \cite{Shi2021Joint, Lee2018Speeding}: 
\vspace{-2mm}
\begin{align}\label{eq:distribution}
    \begin{split}
        P[t_m^k \text{\textless} t]= \left \{
            \begin{array}{ll}
                1-e^{-\frac{\mu_k}{\tau_m D^m_k}(t - \tau_m a_k D^m_k)}, & t\geq \tau_m a_k D^m_k,\\
                0,   & \rm{otherwise},
            \end{array}
        \right.
    \end{split}
\vspace{-2mm}
\end{align}
where the parameters $a_k > 0$ and  $\mu_k > 0$ are the maximum and fluctuation of the computation and communication capability, which is combined into one quantity, of Device $k$, respectively. 
Moreover, we assume that the calculation time of model aggregation has little impact on the training process because of the strong computation capability of the Server and the low complexity of the model. 

The data fairness of Round $r$ corresponding to Job $m$ is indicated by the deviation of the frequency of each device to be scheduled to Job $m$ defined in Formula \ref{eq:fairness}.
\vspace{-1mm}
\begin{equation}\label{eq:fairness}
\mathscr{F}_m^{r}(\mathcal{V}_m^r)=\frac{1}{|\mathcal{K}|}\sum_{k \in \mathcal{K}}(s_{k,m}^r-\frac{1}{|\mathcal{K}|}\sum_{k \in \mathcal{K}}s_{k,m}^r)^2,
\vspace{-2mm}
\end{equation}
where $s_{k,m}^r$ is the frequency of Device $k$ to be scheduled to Job $m$, and $\mathcal{K}$ and $|\mathcal{K}|$ are the set of all devices and the size, respectively. $s_{k,m}^r$ is calculated by counting the total number of the appearance of Device $k$ to be scheduled to Job $m$ in the set of scheduling plans for Job $m$, i.e., $\{\mathcal{V}_m^1,..., \mathcal{V}_m^r\}$.

\subsection{Problem Formulation}

The problem we address is how to reduce the training time when given a loss value for each job. While the execution of each job is carried out in parallel, the problem can be formulated as follows:
\vspace{-1mm}
\begin{align}\label{eq:problem}
    &\displaystyle\min_{\mathcal{V}_m^r}\Big\{ \sum_{m = 1}^M \sum_{r=1}^{R_m'} \mathscr{T}_{m}^{r}(\mathcal{V}_m^r)\Big\}\\
    \text{s.t.} &\begin{cases}
    \displaystyle \mathbb{L}_m(R_m') \leq l_m, \nonumber\\
    \displaystyle\mathcal{V}_m^r \subset \mathcal{K}, \forall m\in \{1,2,...,M\}, \forall r\in \{1,2,...,R_m'\},
    \end{cases}
\vspace{-2mm}
\end{align}
where $l_m$ is the given loss value of Job $m$, $R_m'$ represents the minimum number of rounds to achieve the given loss in the real execution, and $\mathbb{L}_m(R_m')$ is the loss value of the trained model at Round $R_m'$, defined in Formula \ref{eq:eq1}.
As it requires the global information of the whole training process, which is hard to predict, to solve the problem, we transform the problem to the following one, which can be solved with limited local information of each Round. In addition, in order to achieve the given loss value of Job $m$ within a short time (the first constraint in Formula \ref{eq:problem}), we need to consider the data fairness within the total cost in Formula \ref{eq:problem2}, within which the data fairness can help reduce $R_m'$ so as to minimize the total training time.
\vspace{-1mm}
\begin{align}\label{eq:problem2}
    \min_{\mathcal{V}_m^r}\Big\{& TotalCost(\mathcal{V}_m^r)\Big\}, \\
    Total&Cost(\mathcal{V}_m^r) = \sum_{m' = 1}^M Cost_{m'}^{r}(\mathcal{V}_{m'}^r), \label{eq:BatchTotalCost}\\
    s.t. \qquad & \mathcal{V}_{m'}^r \subset \mathcal{K}, \forall {m'}\in \{1,2,...,M\},\nonumber
\vspace{-2mm}
\end{align}
where $Cost_{m}^{r}(\mathcal{V}_m^r)$ can be calculated based on Formula \ref{eq:totalCost} with a set of scheduled devices $\mathcal{V}_m^r$ to be generated using a scheduling method for Job $m$.
Since the scheduling results of one job may have a potential influence on the scheduling of other jobs, we consider the cost of other jobs when scheduling devices to the current job in this problem.
As the search space is $O(2^{|\mathcal{K}|})$, this scheduling problem is still a combinatorial optimization problem \cite{toth2000optimization} and NP-hard \cite{Du1989NPHard, liu2020job}.

\section{Device Scheduling for Multi-job FL}
\label{sec:solutions}

\begin{algorithm}[htbp]		
    \caption{Bayesian Optimization-Based Scheduling}
    \label{alg:bayesian}
    {\bf{Input:}}\\
    \hspace*{0.3in}$\mathcal{V}_{o}:$ A set of occupied devices\\
    \hspace*{0.3in}$S_m:$ A matrix of the frequency of each device sched- \hspace*{0.6in}uled to Job $m$\\
    \hspace*{0.3in}$R_m:$ The maximum round of the current Job $m$\\
    \hspace*{0.3in}$l_m:$ The desired loss value for Job $m$.\\
    {\bf{Output:}}\\
    \hspace*{0.3in} $\mathcal{V}_m = \{\mathcal{V}_{m}^{*1},...,\mathcal{V}_{m}^{*R_m}\}:$ a set of scheduling plans, \hspace*{0.35in}each with the size $|\mathcal{K}| \times C_m$ 
    \begin{algorithmic}[1]
        \State $\Pi_L$ $\leftarrow$ Randomly generate a set of observation points and calculate the cost \label{line:randomGeneration}
        \For{$r \in \{1,...R_m\}$ and $l_m$ is not achieved}
            \State $\Pi'$ $\leftarrow$ Randomly generate a set of observation points \hspace*{0.2in}with the devices within $\mathcal{K} \backslash \mathcal{V}_{o}$ \label{line:randomGenerationInLoop}
            \State $\mathcal{V}_{m}^{*r} \leftarrow \underset{\mathcal{V}\subset \Pi'}{arg\max} \alpha_{\rm EI}(\mathcal{V};\Pi')$ \label{line:chooseNext}
            \State FL training of Job $m$ with $\mathcal{V}_{m}^{*r}$ and update $S_m$, $\mathcal{V}_{o}$ \label{line:boTraining}
            \State $\mathbb{C}_{r}=TotalCost(\mathcal{V}_{m}^{*r})$ \label{line:calculateCost}
            \State $\Pi_{L + r} \gets \Pi_{L + r - 1} \cup (\mathcal{V}_{m}^{*r}, \mathbb{C}_{r})$\label{line:addPoint}
        \EndFor
   \end{algorithmic}
\end{algorithm}
In this section, we propose two scheduling methods, i.e., BO-based and RL-based, to address the problem defined in Formula \ref{eq:problem2}. The scheduling plan generated by a scheduling method is defined in Formula \ref{eq:schedulingPlan}:
\vspace{-2mm}
\begin{align}\label{eq:schedulingPlan}
    \mathcal{V'}_m^r = \argmin_{\mathcal{V}_m^r\subset {\{\mathcal{K} \backslash \mathcal{V}_{o}^r\}}} TotalCost(\mathcal{V}_m^r),
\vspace{-2mm}
\end{align}
where $\mathcal{V'}_m^r$ is a scheduling plan, $\mathcal{K} \backslash \mathcal{V}_{o}^r$ represents the set of available devices to schedule, $TotalCost(\mathcal{V}_m^r)$ is defined in Formula \ref{eq:BatchTotalCost}, and $\mathcal{K}$ and $\mathcal{V}_{o}^r$ are the set of all devices and the set of occupied devices in Round $r$, respectively. 

\subsection{Bayesian Optimization-Based Scheduling}

While the Gaussian Process (GP) \cite{Srinivas2010Gaussian} can well represent linear and non-linear functions, BO-based methods \cite{Shahriari2061BO} can exploit a GP to find a near-optimal solution for the problem defined in Formula \ref{eq:schedulingPlan}.
In this section, we propose a Bayesian Optimization-based Device Scheduling method (BODS).

We adjust a GP to fit the cost function $TotalCost(\cdot)$. The GP is composed of a mean function $\mu$ defined in Formula \ref{eq:mu} and a covariance function $\rm{K}$ defined in Formula \ref{eq:kernel} with a $Matern$ kernel \cite{williams2006Ggaussian}.
\begin{align}\label{eq:mu}
\mu(\mathcal{V}_{m}^r)=\underset{\mathcal{V}_{m}^r \subset {\{\mathcal{K} \backslash \mathcal{V}_{o}^r\}}}{\mathbb{E}}[TotalCost(\mathcal{V}_{m}^r)]
\end{align}
\begin{align}\label{eq:kernel}
& {\rm K}(\mathcal{V}_{m}^r, \mathcal{V}_{m}'^r) = \underset{\mathcal{V}_{m}^r \subset {\{\mathcal{K} \backslash \mathcal{V}_{o}^r\}}, \mathcal{V}_{m}'^r \subset {\{\mathcal{K} \backslash \mathcal{V}_{o}^r\}}}{\mathbb{E}}\nonumber\\
& [(TotalCost(\mathcal{V}_{m}^r)-\mu(\mathcal{V}_{m}^r)) (TotalCost(\mathcal{V}_{m}'^r)-\mu(\mathcal{V}_{m}'^r))]
\end{align}

The BODS is explained in {\bf{Algorithm \ref{alg:bayesian}}}.
First, we randomly generate a set of observation points and calculate the cost based on Formula \ref{eq:totalCost} (Line \ref{line:randomGeneration}). 
Each observation point is a pair of scheduling plan and cost for the estimation of mean function and the covariance function. Then, within each round, we randomly sample a set of scheduling plans (Line \ref{line:randomGenerationInLoop}), within which we select the one with the biggest reward using updated $\mu$ and $K$ based on $\Pi_{L + r - 1}$ (Line \ref{line:chooseNext}). Afterward, we perform the FL training for Job $m$ with the generated scheduling plan (Line \ref{line:boTraining}), and calculate the cost corresponding to the real execution (Line \ref{line:calculateCost}) according to Formula \ref{eq:BatchTotalCost} and update the observation point set (Line \ref{line:addPoint}).

\begin{figure}[htbp]
\centering
\centerline{\includegraphics[width=1.0\linewidth]{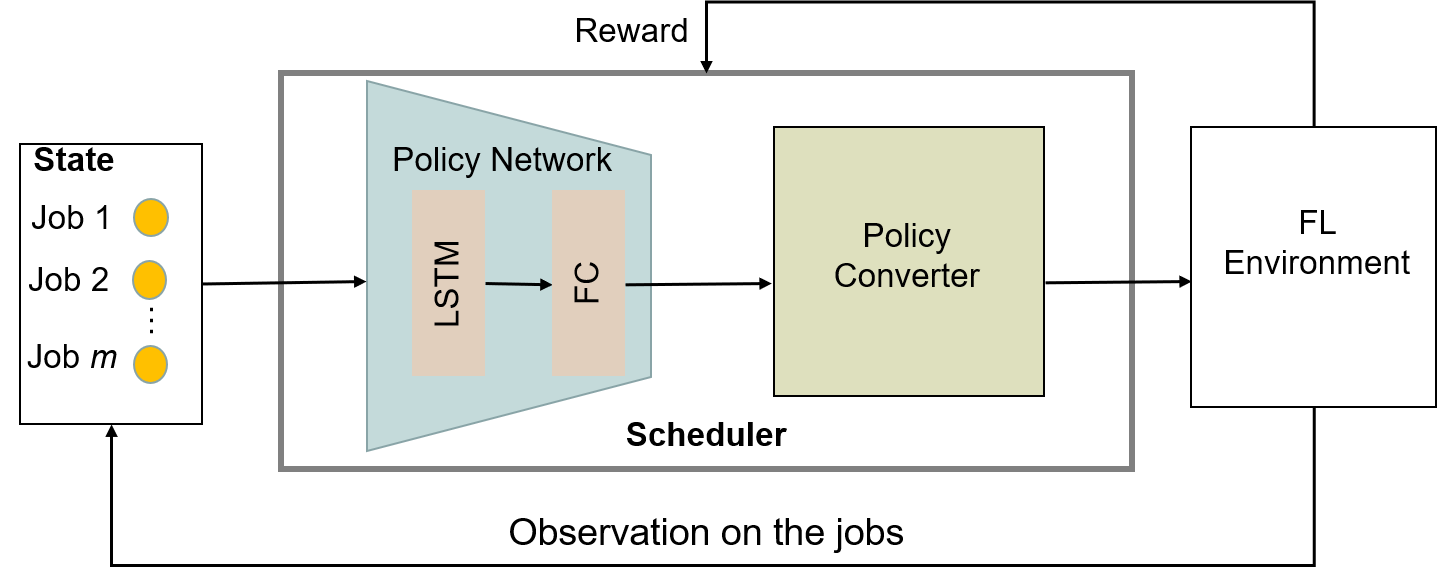}}
\vspace{-2mm}
\caption{The architecture of the RLDS.}
\label{RLDS-framework}
\vspace{-8mm}
\end{figure}

Let ($\mathcal{V}_l^r$, $\mathbb{C}_l$) denote an observation point $l$ for Job $m$ in Round $r$, where $\mathcal{V}_l^r=\{\mathcal{V}_{l, 1}^r,...,\mathcal{V}_{l, M}^r\}$ and $\mathbb{C}_l$ is the cost value of $TotalCost(\mathcal{V}_{l,m}^r)$ while the scheduling plans of other jobs are updated with the ones in use in Round $r$.
At a given time, we have a set of observations $\Pi_{L-1}=\{(\mathcal{V}_1^r, \mathbb{C}_1),...,(\mathcal{V}_{L-1}^r, \mathbb{C}_{L-1})\}$ composed of $L-1$ observation points.
We denote the minimum cost value within the $L - 1$ observations by $\mathbb{C}^{+}_{L-1}$. Then, we exploit Expected Improvement(EI) \cite{Jones1998Efficient} to choose a new scheduling plan $\mathcal{V}_{m}^{*r}$ in Round $r$ that improves $\mathbb{C}^{+}_{L-1}$ the most, which is the utility function.
Please note that this is not an exhaustive search as we randomly select several observation points (a subset of the whole search space) at the beginning and add new observation points using the EI method.

\subsection{Reinforcement Learning-Based Scheduling}
In order to learn more information about the near-optimal scheduling patterns for complex jobs, we further propose a Reinforcement Learning-based Device Scheduling (RLDS) method as shown in Fig. \ref{RLDS-framework}, which is inspired by \cite{Hongzi2019rniLeang, sun2020deepweave}. The scheduler of RLDS consists of a policy network and a policy converter. In the process of device scheduling, RLDS collects the status information of jobs as the input to the policy network. Then, the policy network generates a list of probabilities on all devices as the output. Finally, the policy converter converts the list into a scheduling plan.

\subsubsection{Policy Network} The policy network is implemented using a Long Short-Term Memory (LSTM) network followed by a fully connected layer, which can learn the sharing relationship of devices among diverse jobs. We take the computation and communication capability of available devices to be used in Formula \ref{eq:distribution}, and the data fairness of each job defined in Formula \ref{eq:fairness} as the input. The network calculates the probability of each available device to be scheduled to a job.

\begin{figure*}[htbp]
\centering
\begin{subfigure}{0.3\linewidth}
\includegraphics[width=\linewidth]{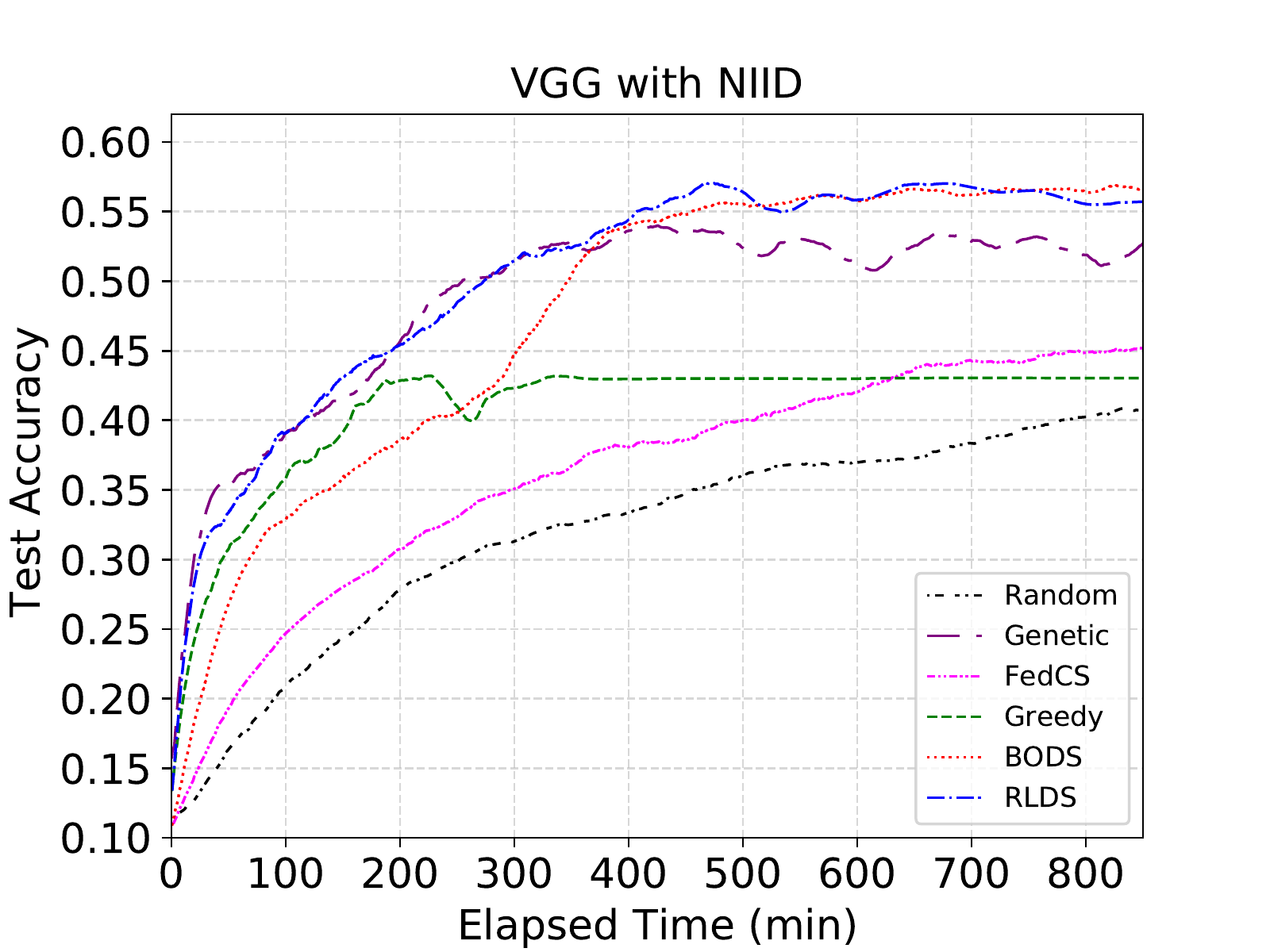}
\caption{}
\label{figVgNiid}
\end{subfigure}
\begin{subfigure}{0.3\linewidth}
\includegraphics[width=\linewidth]{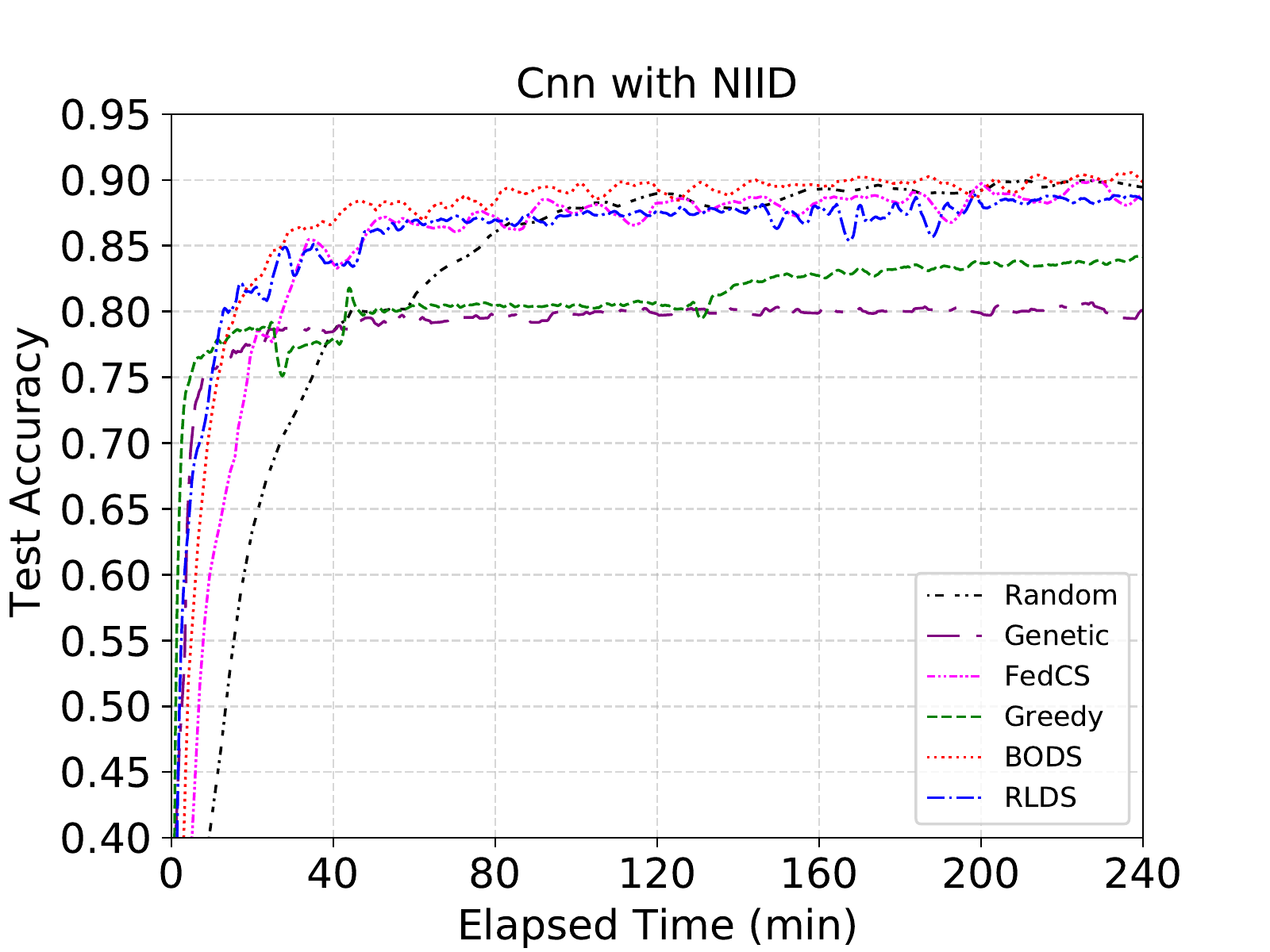}
\caption{}
\label{figCNiid}
\end{subfigure}
\begin{subfigure}{0.3\linewidth}
\includegraphics[width=\linewidth]{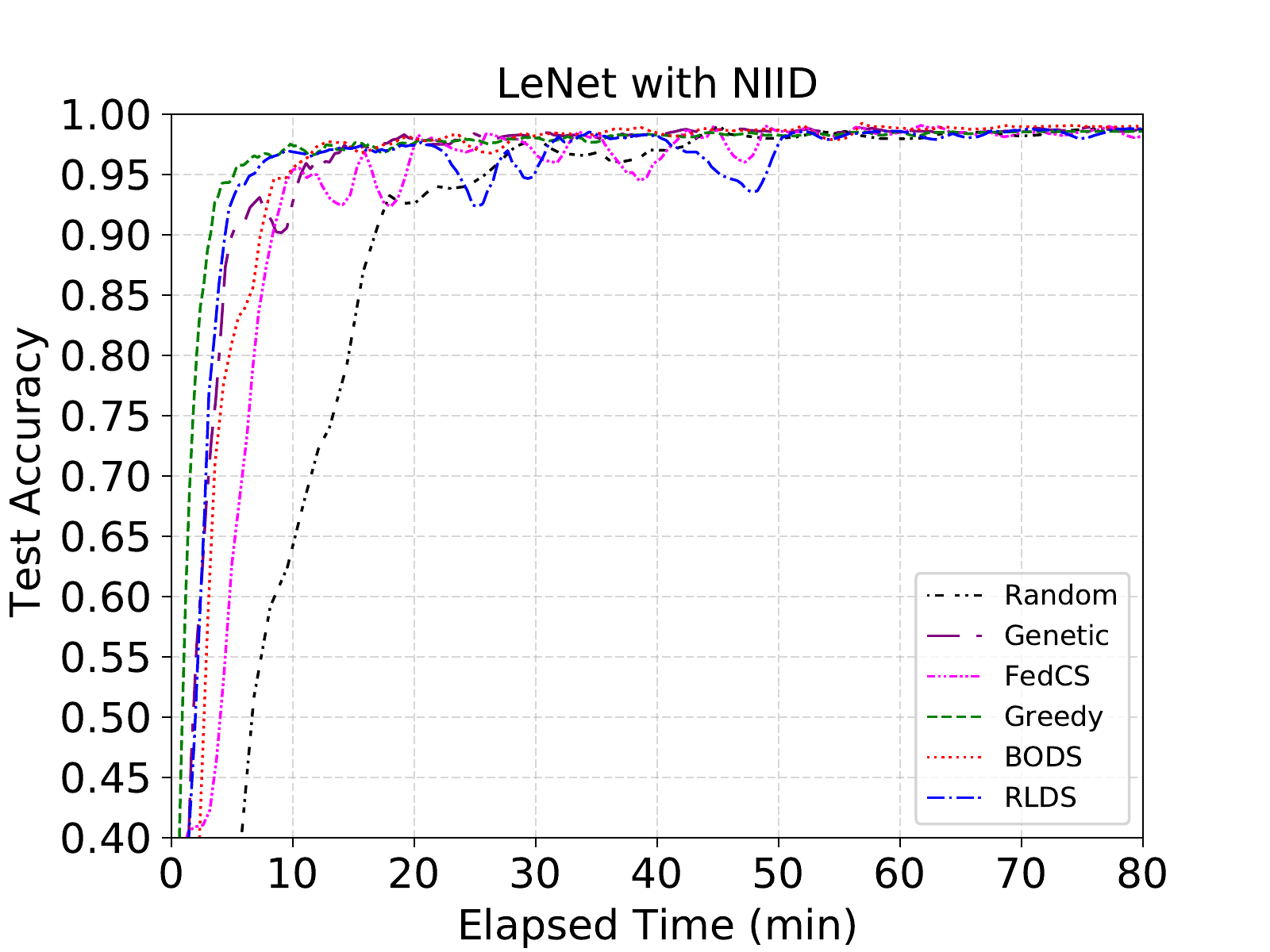}
\caption{}
\label{figLeNiid}
\end{subfigure}
\vspace{-2mm}
\caption{The accuracy of different jobs in Group A over time with the non-IID distribution.}
\label{fig:accuracy}
\vspace{-4mm}
\end{figure*}

\begin{algorithm}[ht]		
    \caption{Reinforcement Learning-Based Scheduling}
    \label{alg:reinforcement}
    {\bf{Input:}}\\
    \hspace*{0.3in}$\mathcal{V}_{o}:$ A set of occupied devices\\
    \hspace*{0.3in}$S_m:$ A vector of the frequency of each device sched- \hspace*{0.6in}uled to Job $m$\\
    \hspace*{0.3in}$R_m:$ The maximum round of the current Job $m$\\
    \hspace*{0.3in}$l_m:$ The desired loss value for Job $m$.\\
    {\bf{Output:}}\\
    \hspace*{0.3in} $\mathcal{V}_m = \{\mathcal{V}_{m}^{1},...,\mathcal{V}_{m}^{R_m}\}:$ a set of scheduling plans, \hspace*{0.35in}each with the size $|\mathcal{K}| \times C_m$ 
    
    \begin{algorithmic}[1]
        \State $\theta$ $\leftarrow$ pre-trained policy network, $\Delta\theta\leftarrow0$, $b_m\leftarrow0$ \label{line:loadModel}
        \For{$r \in \{1,2,...,R_m\}$ and $l_m$ is not achieved}
            \State $\mathcal{V}_{m}^{r} \leftarrow$ generate a scheduling plan using the policy \hspace*{0.2in}network \label{line:generatePlan}
            \State FL training of Job $m$ and update $S_m$, $\mathcal{V}_{o}$ \label{line:training}
            \State Compute $\mathscr{R}^m$\label{line:computeRD}
            \State Update $\theta$ according to Formula \ref{eq:rlupdate} \label{line:updateTheta}
            \State $b_m$ $\leftarrow$ (1 - $\gamma)$ * $b_m$ + $\gamma$ * $\mathscr{R}_n^m$ \label{line:updateBm}
        \EndFor
   \end{algorithmic}
\end{algorithm}

\subsubsection{Policy Converter} The Policy Converter generates a scheduling plan based on the probability of each available device calculated by the policy network with the $\epsilon$-greedy strategy \cite{xia2015online}.

\subsubsection{Training} In the training process of RLDS, we define the reward as $\mathscr{R}^m = -1 * TotalCost(\mathcal{V}_m^r)$. Inspired by \cite{williams1992simple, Zoph2017Neural}, we exploit Formula \ref{eq:rlupdate} to update the policy network:
\vspace{-4mm}
\begin{equation}\label{eq:rlupdate}
\begin{aligned}
    \theta^{'} = \theta + &\frac{\eta}{N} \sum_{n=1}^N \sum_{k \in \mathcal{V}_{n,m}^{r}}^{ \mathcal{V}_{n,m}^{r} \subset \mathcal{K}\backslash\mathcal{V}_{o}^r} \nabla_\theta &\log P(\mathscr{S}_k^m|\mathscr{S}_{(k-1):1}^m; \theta) \\&(\mathscr{R}_n^m - b_m),
\end{aligned}
\end{equation}
where $\theta'$ and $\theta$ represent the updated parameters and the current parameters of the policy network, respectively, $\eta$ is the learning rate, $N$ is the number of scheduling plans to update the model in Round $r$ ($N > 1$ in the pre-training process and $N = 1$ during the execution of multiple jobs), $P$ represents the probability calculated based on the RL model, $\mathscr{S}_k^m = 1$ represents that Device $k$ is scheduled to Job $m$, and $b_m$ is the baseline value for reducing the variance of the gradient. 

We exploit RLDS during the training process of multiple jobs within the MJ-FL framework as shown in {\bf{Algorithm \ref{alg:reinforcement}}}. We pre-train the policy network with randomly generated scheduling plans (see details in Appendix) (Line \ref{line:loadModel}). 
When generating a scheduling plan for Job $m$, the latest policy network is utilized (Line \ref{line:generatePlan}). We perform the FL training for Job $m$ with the generated scheduling plan and update the frequency matrix $S_m$ and the set of occupied devices $\mathcal{V}_{o}$ (Line \ref{line:training}). Afterward, we calculate the reward corresponding to the real execution (Line \ref{line:computeRD}). The parameters are updated based on the Formula \ref{eq:rlupdate} (Line \ref{line:updateTheta}), while the baseline value $b_m$ is updated while considering the historical value (Line \ref{line:updateBm}).

\begin{table*}[ht]
  \caption{The convergence accuracy and the time required to achieve the target accuracy for different methods in Group A. The numbers in parentheses represent the target accuracy, and "/" represents that the target accuracy is not achieved.}
  \setlength\tabcolsep{2.5pt}
  \label{tab:GroupA}
  \begin{tabular}{ccccccc ccccccc}
    \toprule
    &\multicolumn{6}{c}{Convergence Accuracy} & &\multicolumn{6}{c}{Time (min)}\\
    \cline{2-7} \cline{9-14}
    &  Random& Genetic& FedCS& Greedy& BODS& RLDS  & & Random& Genetic& FedCS& Greedy& BODS& RLDS\\
    \cline{1-14}
    &\multicolumn{13}{c}{Non-IID}\\
    
    \midrule
    VGG& 0.55& 0.54& 0.55& 0.43& \bf{0.57}& \bf{0.57}&  
    VGG(0.55)& 2486& 1164.3& 1498.5& /& 455.1& \bf{406.8}\\
    CNN& \bf{0.90}& 0.80& 0.80& 0.83& \bf{0.90}& 0.88&  
    CNN(0.80)& 44.25& 95.85& 27.39& 43.04& 15.88& \bf{12.75}\\
    LeNet& 0.990& 0.988& 0.990& 0.986& \bf{0.991}& 0.990&  LeNet(0.984)& 43.81& 30.15& 33.37& 43.76& \bf{28.93}& 34.08\\
    
     \cline{1-14}
    &\multicolumn{13}{c}{IID}\\
    \cline{1-14}
    VGG& {\bf{0.614}}& 0.558& 0.603& 0.522& 0.603& \bf{0.614}&  VGG(0.60)& 529.9& /& 322.5& /& 293.6& {\bf{249.2}}\\
    CNN& \bf{0.943}& 0.928& \bf{0.943}& 0.928& \bf{0.943}& 0.937&  CNN(0.930)& 52.05& 176.85& 27.45& 26.48& 19.25& \bf{18.29}\\
    LeNet& 0.9945& 0.9928& 0.9934& 0.99& \bf{0.9946}& 0.9933&  LeNet(0.993)& 43.15& 57.53& 27.31& /& \bf{16.73}& 23.31\\
    
  \bottomrule
\end{tabular}
\end{table*}

\begin{table*}[tb]
  \caption{The convergence accuracy and the time required to achieve the target accuracy for different methods in Group B. The numbers in parentheses represent the target accuracy, and "/" represents that the target accuracy is not achieved.}
  \setlength\tabcolsep{1.8pt}
  \label{tab:GroupB}
  \begin{tabular}{ccccccc ccccccc}
    \toprule
     &\multicolumn{6}{c}{Convergence Accuracy} & &\multicolumn{6}{c}{Time (min)}\\
    \cline{2-7} \cline{9-14}
    &  Random& Genetic& FedCS& Greedy& BODS& RLDS  & & Random& Genetic& FedCS& Greedy& BODS& RLDS\\
    \cline{1-14}
    &\multicolumn{13}{c}{Non-IID}\\
    
    \midrule
    ResNet& 0.546& 0.489& 0.523& 0.403& \bf{0.583}& 0.537&  ResNet(0.45)& 571.0& 307.2& 279.5& 174.2& 157.5& \bf{137.6}\\
    CNN& 0.821& 0.767& 0.821& 0.764& \bf{0.836}& 0.823&  CNN(0.73)& 47.1& 22.0& 18.5& 70.8& 13.8& \bf{4.8}\\
    AlexNet& 0.989& 0.986& 0.987& 0.871& \bf{0.990}& 0.989&  AlexNet(0.978)& 141.85& 77.74& 84.8& /& 61.91& \bf{57.97}\\
    
    \cline{1-14}
    &\multicolumn{13}{c}{IID}\\
    \cline{1-14}
    ResNet& 0.787& 0.754& 0.782& 0.743& \bf{0.791}& 0.771&  ResNet(0.740)& 65.93& 32.51& 31.4& 52.93& 15.9& \bf{11.96}\\
    CNN& 0.867& 0.867& 0.868& 0.868& \bf{0.869}& \bf{0.869}&  CNN(0.865)& 88.81& 23.89& 26.06& 21.42& 23.99& \bf{9.3}\\
    AlexNet& 0.9938& 0.9938& 0.9939& 0.9935& 0.9939& \bf{0.9943}&  AlexNet(0.9933)& 35.08& 19.44& 20.97& /& 21.65& \bf{12.58}\\
  \bottomrule
\end{tabular}
\vspace{-4mm}
\end{table*}

\section{Experiments}
\label{sec:experiment}
In this section, we present the experimental results to show the efficiency of our proposed scheduling methods within MJ-FL. We compared the performance of RLDS and BODS with four baseline methods, i.e., Random \cite{McMahan2017Communication-efficien}, FedCS \cite{Nishio2019Client}, Genetic \cite{barika2019scheduling}, and Greedy \cite{shi2020device}. 

\subsection{Federated Learning Setups}

In the experiment, we take three jobs as a group to be executed in parallel. 
We carry out the experiments with two groups, i.e., Group A with VGG-16 (VGG) \cite{simonyan2015very}, CNN (CNN-A-IID and CNN-A-non-IID) \cite{lecun1998gradient}, and LeNet-5 (LeNet) \cite{lecun1998gradient}, and Group B with Resnet-18 (ResNet) \cite{He2016}, CNN (CNN-B) \cite{lecun1998gradient}, and Alexnet \cite{Krizhevsky2012ImageNet}, while each model corresponds to one job. The complexity of the models is as follows: AlexNet $<$ CNN-B $<$ ResNet and LeNet $<$ CNN (CNN-A-IID and CNN-A-non-IID) $<$ VGG. We exploit the datasets of CIFAR-10 \cite{krizhevsky2009learning}, emnist-letters \cite{cohen2017emnist}, emnist-digital \cite{cohen2017emnist}, Fashion-MNIST \cite{xiao2017fashion}, and MNIST \cite{lecun1998gradient} in the training process. Please see details of the models and datasets in Appendix. 
For the non-IID setting of each dataset, the training set is classified by category, and the samples of each category are divided into 20 parts. Each device randomly selects two categories and then selects one part from each category to form its local training set. For the IID setting, each device randomly samples a specified number of images from each training set. 
In addition, we use 12 Tesla V100 GPUs to simulate an FL environment composed of a parameter server and 100 devices. We use Formula \ref{eq:distribution} to simulate the capabilities of devices in terms of training time with the uniform sampling strategy, while the accuracy is the results from the real training processes. In the experimentation, we use corresponding target accuracy (for ease of comparison) in the place of target loss value.

{\bf{Evaluation on the non-IID setting:}} When the decentralized data is of non-IID, the data fairness defined in Formula \ref{eq:fairness} has a significant influence on the accuracy. As shown in Fig. \ref{fig:accuracy}, the convergence speed of our proposed methods, i.e., RLDS and BODS, is significantly faster than other methods. RLDS has a significant advantage for complex jobs (VGG in Fig. \ref{figVgNiid}), while BODS can lead to good performance for relatively simple jobs in Groups A and B (please see details of Group B in Fig. \ref{fig:groupBNonIID} in Appendix). 
In addition, as shown in Tables \ref{tab:GroupA} and \ref{tab:GroupB}, the final accuracy of RLDS and BODS outperforms other methods (up to 44.6\% for BODS and 33.3\% for RLDS), as well. Given a target accuracy, our proposed methods can achieve the accuracy within a shorter time, compared with baseline methods, in terms of the time for a single job, i.e., the training time of each job (up to 5.04 times shorter for BODS and 5.11 times shorter for RLDS), and the time for the whole training process, i.e., the total time calculated based on Formula \ref{eq:problem} (up to 4.15 times for BODS and 4.67 times for RLDS), for Groups A and B. We have similar observations with IID, while the advantage of RLDS is much more significant (up to 8.67 times shorter in terms of the time for a single job) than that of non-IID as shown in Tables \ref{tab:GroupA} and \ref{tab:GroupB}. We also find that MJ-FL outperforms (up to 5.36 faster and 12.5\% higher accuracy) sequential execution of single-job FL (see details in Appendix).

As RLDS can learn more information with a complex neural network, RLDS outperforms BODS for complex jobs. BODS can lead to high convergence accuracy and fast convergence speed thanks to the emphasis on the combination of the data fairness and the capability of the device, i.e., computation and communication capability. Both RLDS and BODS significantly outperform the baseline methods, while there are also differences among the four methods. The Greedy method is more inclined to schedule the devices with high capability, which leads to a significant decrease in the final convergence accuracy. The Genetic method can exploit randomness to achieve data fairness while generating scheduling plans, and the convergence performance is better than the Greedy method. The FedCS method optimizes the scheduling plan with random selection, which improves the fairness of the device to a certain extent, and the convergence speed is faster than the Random method.

\section{Conclusion}
\label{sec:conclusion}

In this work, we proposed a new Multi-Job Federated Learning framework, i.e., MJ-FL. The framework is composed of a system model and two device scheduling methods. The system model is composed of a process for the parallel execution of multiple jobs and a cost model based on the capability of devices and data fairness. We proposed two device scheduling methods, i.e., RLDS for complex jobs and BODS for simple jobs, to efficiently select proper devices for each job based on the cost model. We carried out extensive experimentation with six real-life models and four datasets with IID and non-IID distribution. The experimental results show that MJ-FL outperforms the single-job FL, and that our proposed scheduling methods significantly outperform baseline methods (up to 44.6\% in terms of accuracy, 8.67 times faster for a single job, and 4.67 times faster for the total time). 

\newpage

\bibliography{IEEEabrv}

\newpage

\section*{Appendix}

\begin{algorithm}[ht]		
    \caption{Reinforcement Learning Based Pre-Training}
    \label{alg:reinforcement-pre-training}
    {\bf{Input:}}\\
    \hspace*{0.3in}$\mathcal{V}_{o}:$ A set of occupied devices\\
    \hspace*{0.3in}$S_m:$ A vector of the frequency of each device sched- \hspace*{0.6in}uled to Job $m$\\
    \hspace*{0.3in}$N:$ The number of scheduling plans used to train the \hspace*{0.55in}network for each round\\
    \hspace*{0.3in}$R_m:$ The maximum round of the current Job $m$\\
    \hspace*{0.3in}$l_m:$ The desired loss value for Job $m$.\\
    {\bf{Output:}}\\
    \hspace*{0.3in} $\theta:$ Parameters of the pre-trained policy network
    
    \begin{algorithmic}[1]
        \State $\theta$ $\leftarrow$ randomly initialize the policy network, $b_m\leftarrow0$ \label{line:ap:loadModel}
        \For{$r \in \{1,2,...,R_m\}$ and $l_m$ is not achieved}
            \State $\mathcal{V}_{m}^{r} \leftarrow$ generate a set of $N$ scheduling plans\label{line:ap:generatePlan}
            \For{$\mathcal{V}_{n, m}^{r} \in \mathcal{V}_{m}^{r}$}
                \State $Rd_n^m \leftarrow$ TotalCost($\mathcal{V}_{n,m}^{r}$) \label{line:ap:computeRD}
            \EndFor
            \State Update $\theta$ according to Formula \ref{eq:rlupdate} \label{line:ap:updateTheta}
            \State $b_m$ $\leftarrow$ (1 - $\gamma)$ * $b_m$ + $\frac{\gamma}{N}$ * $\sum_{n=1}^N \mathscr{R}_n^m$ \label{line:ap:updateBm}
            \State $\mathcal{V}_{m}^{*r}$ $\leftarrow$ $\argmin_{\mathcal{V}_{n,m}^{r} \in \mathcal{V}_{m}^{r} }$TotalCost($\mathcal{V}_{n,m}^{r}$)\label{line:ap:bestSchedulingPlan}
            \State Update $S_m$, $\mathcal{V}_{o}$ with $\mathcal{V}_{m}^{*r}$\label{line:ap:updateParameters}
        \EndFor
   \end{algorithmic}
\end{algorithm}

\subsection{Loss Estimation}
We assume exploiting stochastic gradient descent (SGD) to train models, which converges at a rate of $O(r)$ with $r$ representing the number of rounds \cite{2018Optimus}. Inspired by \cite{2019On}, we exploit Formula \ref{eq:accuracy} to roughly estimate the loss value of the global model for Job $m$ at Round $r$.
\begin{equation}\label{eq:accuracy}
Loss_m(r) = \frac{1}{\beta^0_m r + \beta^1_m} + \beta^2_m,
\end{equation}
where $\beta_m^0$, $\beta_m^1$ and $\beta_m^2$ represent non-negative coefficients of the convergence curve of Job $m$. $\beta_m^0$, $\beta_m^1$ and $\beta_m^2$ can be calculated based on previous execution.
In addition, we assume that the real number of rounds corresponding to the same loss value has 30\% error compared with $r$ (from the observation of multiple execution). Given a loss value of a model, we use this loss estimation method to calculate the maximum rounds for each job. Given a loss value of a model, we use this loss estimation method to calculate the number of rounds as $R_m^c$ and use $(1+0.3) * R_m^c$ as $R_m$ defined in Table \ref{tab:summary}. Please note that this estimation is different from the loss value during the real execution; i.e., $R_m'$ can be different from $R_m$.

\subsection{Details for Bayesian Optimization-Based Scheduling}

\begin{table}[htbp]
  \caption{Experimental Setup of Group A. Size represents the size of training samples and test samples (number of training samples/number of test samples). ``Emnist-L'' represents ``Emnist-Letters'' and ``Emnist-D'' represents ``Emnist-Digitals''.}
  \label{tab:setup-GroupA}
  \begin{tabular}{cccl}
    \toprule
    datasets&  Cifar10& Emnist-L & Emnist-D\\
    \midrule
    Features& 32x32&  28x28& 28x28\\
    Network model& VGG16& CNN& LeNet5\\
    Parameters& 26,233K& 3,785K& 62K\\
    Size & 50k/10k & 124.8k/20.8k & 240k/40k\\
    Local epochs& 5& 5& 5\\
    Mini-batch size& 30& 10& 64\\
  \bottomrule
\end{tabular}
\end{table}
\begin{table}[htbp]
  \caption{Experimental Setup of Group B. Size represents the size of training samples and test samples (number of training samples/number of test samples).}
  \setlength\tabcolsep{5pt}
  \label{tab:setup-GroupB}
  \begin{tabular}{cccl}
    \toprule
    datasets&   Fashion\_{mnist}&  Cifar10&  Mnist\\
    \midrule
    Features& 28x28&  32x32& 28x28\\
    Network model& CNN& ResNet18& AlexNet\\
    Parameters& 225K& 598K& 3,275K\\
    Size & 60K/10K & 50K/10K & 60K/10K\\
    Local epochs& 5& 5& 5\\
    Mini-batch size& 10& 30& 64\\
  \bottomrule
\end{tabular}
\end{table}
The utility function is defined in Formula \ref{eq:equility}.
\begin{equation}\label{eq:equility}
    u(\mathcal{V}_{m}^{*r})=max(0, \mathbb{C}^{+}_{L-1}-TotalCost(\mathcal{V}_{m}^{*r})),
\end{equation}
where we receive a reward $\mathbb{C}^{+}_{L-1}-TotalCost(\mathcal{V}_{m}^{*r})$ if $TotalCost(\mathcal{V}_{m}^{*r})$ turns out to be less than $\mathbb{C}^{+}_{L-1}$, and no reward otherwise. Then, we use the following formula, which is also denoted an acquisition function, to calculate the expected reward of a given scheduling plan $\mathcal{V}$.
\begin{equation}\label{eq:eq7}
\begin{aligned}
    \alpha_{\rm EI}(\mathcal{V};\Pi_{L-1}) = &\mathbb{E}[u(\mathcal{V})|\mathcal{V}, \Pi_{L-1}]\\
    =&(\mathbb{C}^{+}_{L-1}-\mu(\mathcal{V}))\Phi(\mathbb{C}^{+}_{L-1}; \mu(\mathcal{V}), 
    {\rm K}(\mathcal{V},\mathcal{V}))\\
    &+ {\rm K}(\mathcal{V},\mathcal{V})\mathcal{N}(\mathbb{C}^{+}_{L-1}; \mu(\mathcal{V}), \rm K(\mathcal{V},\mathcal{V})),
 \end{aligned}
\end{equation}
where $\Phi$ is the Cumulative Distribution Function (CDF) of the standard Gaussian distribution. Finally, we can choose the scheduling plan with the largest reward as the next observation point, i.e., $\mathcal{V}_{L,m}^{*r}$.

\subsection{Training Process of Reinforcement Learning-Based Device Scheduling}

We pre-train the policy network using {\bf{Algorithm \ref{alg:reinforcement-pre-training}}}. First, we randomly initialize the policy network (Line \ref{line:ap:loadModel}). We use the latest policy network and the $\epsilon$-Greedy method to generate $N$ scheduling plans (Line \ref{line:ap:computeRD}). The parameters are updated based on the Formula \ref{eq:rlupdate} (Line \ref{line:ap:updateTheta}), and the baseline value $b_m$ is also updated with the consideration of the historical value (Line \ref{line:ap:updateBm}). Afterward, we choose the best scheduling plan that corresponds to the minimum total cost, i.e., the maximum reward (Line \ref{line:ap:bestSchedulingPlan}). Finally, we update the frequency matrix $S_m$ and the set of occupied devices $\mathcal{V}_{o}$, while assuming that the best scheduling plan is used for the multi-job FL (Line \ref{line:ap:updateParameters}).

\subsection{Details of Experimental Setup}

\begin{table*}[htbp]
  \caption{The time required to achieve the target accuracy for jobs executed sequentially with FedAvg. "*" indicates that it fails to achieve the target accuracy.}
  \setlength\tabcolsep{10pt}
  \renewcommand{\arraystretch}{1.3}
  \label{tab:Signal}
  \begin{tabular}{cccc cccc}
    \toprule
     & \multicolumn{3}{c}{NIID/IID}&  & \multicolumn{3}{c}{NIID/IID}\\
    \cline{2-4} \cline{6-8}
    Job&  VGG& CNN& LeNet& & ResNet& CNN& AlexNet\\
    \cline{1-8}
    Target Accuracy&  0.55/0.60& 0.80/0.93& 0.984/0.993& &
       0.45/0.74& 0.73/0.865& 0.978/0.9933\\
    \midrule
    Time (min)&  2483.4/414.6& 53.1/45.5& 50.5/52.1&  &
               594.3/*& 36.1/172.9& 127.3/65.16\\
 \bottomrule
 \end{tabular}
\end{table*}
\begin{table*}[htbp]
\caption{Summary of Main Notations}
\begin{center}
\begin{tabular}{cc}
\toprule
Notation & Definition \\
\hline
\label{tab:summary}
$\mathcal{K}$; $|\mathcal{K}|$ & Set of all devices; size of $\mathcal{K}$ \\
$M$; $m$; $T$ & The total number of jobs; index of jobs; total training time \\
$\mathcal{D}_k^m$; $D_k^m$; $d_k^m$ & Local dataset of Job $m$ on Device $k$; size of $\mathcal{D}_k^m$; batch size of the local update of Device $k$ \\
$\mathcal{D}^m$; $D^m$ & Global dataset of Job $m$; size of $\mathcal{D}^m$ \\
$F_k^m(\boldsymbol{w})$; $F^m(\boldsymbol{w})$ & Local loss function of Job $m$ in Device $k$; global loss function of job $m$ \\
$\boldsymbol{w}_{k,r}^m(j)$ & Local model of Device $k$ in the $j$-th local update of Round $r$ \\
$R_m$ & The maximum rounds for job $m$ during the execution \\
$R_m'$ & The maximum rounds for job $m$ to achieve the required performance (loss value or accuracy) \\
$l_m$ & The desired loss value for job $m$\\
$\tau_m$; $C_m$ &Number of local epochs of Job $m$; the ratio between the number of devices scheduled to Job $m$ and $|\mathcal{K}|$\\
$S_m, s_{k,m}^r$ & The frequency vector for Job $m$; the frequency of Device $k$ scheduled to Job $m$ at Round $r$\\
$\mathcal{V}_m^r$ & A set of devices scheduled to Job $m$ at Round $r$\\
$\mathcal{V}_{o}; \mathcal{V}_{o}^r$ & A set of occupied devices; the set of occupied devices in Round $r$ \\
\bottomrule
\end{tabular}
\end{center}
\end{table*}
CNN-A-IID is composed of two $3\times 3$ convolution layers, one with 32 channels and the other with 64 channels. Each layer is followed by one batch normalization layer and $2\times 2$ max pooling. Then, after the two convolution layers, there are one flatten layer and three fully-connected layers (1568, 784, and 26 units). Since the convergence behavior of CNN on non-IID in Group A is not good, we make a simple modification of CNN-A-IID to CNN-A-non-IID. CNN-A-non-IID consists of three $3\times 3$ convolution layers (32, 64, 64 channels, each of them exploits ReLU activations, and each of the first two convolution layers is followed by $2\times 2$  max pooling), followed by one flatten layer and two fully-connected layers (64, 26 units). CNN-B consists of two $2\times 2$ convolution layers (64, 32 channels, each of them exploits ReLU activations) followed by a flatten layer and a fully-connected layer, and each convolution layer is followed by a dropout layer with 0.05. In addition, the other parameters are shown in Tables \ref{tab:setup-GroupA} and \ref{tab:setup-GroupB}. 

\subsection{Explanation of Notations}

The meanings of the major notations in this paper are summarized in Table \ref{tab:summary}. In particular, $S_m = \{s_{1, m}, ..., s_{|\mathcal{K}|,m}\}$ represents the frequency vector of Job $m$. At the beginning, i.e., Round 0, each $s_{k, m} \in S_m$ is 0. Let $s_{k, m}^r$ represent the frequency of Device $k$ scheduled to Job $m$ at Round $r$. Then, we can calculate $s_{k, m}^{r+1}$ using the following formula:
\begin{align}
s_{k, m}^{r+1} = 
\begin{cases}
\displaystyle s_{k, m}^{r} + 1,\hspace*{0.45in}$if$~$Device$~k \in \mathcal{V}_m^r \\
\displaystyle s_{k, m}^{r},\hspace*{0.7in}$otherwise$
\end{cases}
\end{align}

\subsection{Comparison With Single-Job Federated Learning}

In order to prove the effectiveness of our proposed framework, i.e., MJ-FL, over the Single-Job FL (SJ-FL) approach, we executed each group of jobs sequentially with FedAvg, which is denoted the Random method when adapted to multi-job FL. As shown in Table \ref{tab:Signal}, RLDS with MJ-FL outperforms the FedAvg with SJ-FL, which executes jobs sequentially, up to 5.36 times faster in terms of the training time while achieving the same accuracy. Similarly, the advantage of BODS can be up to 4.68 times faster. In addition, Random within MJ-FL also outperforms SJ-FL up to 0.25 times faster.

\subsection{Comparison With Multiple Targets and Other Methods}

\begin{figure*}[htbp]
\centering
\begin{subfigure}{0.3\linewidth}
\includegraphics[width=\linewidth]{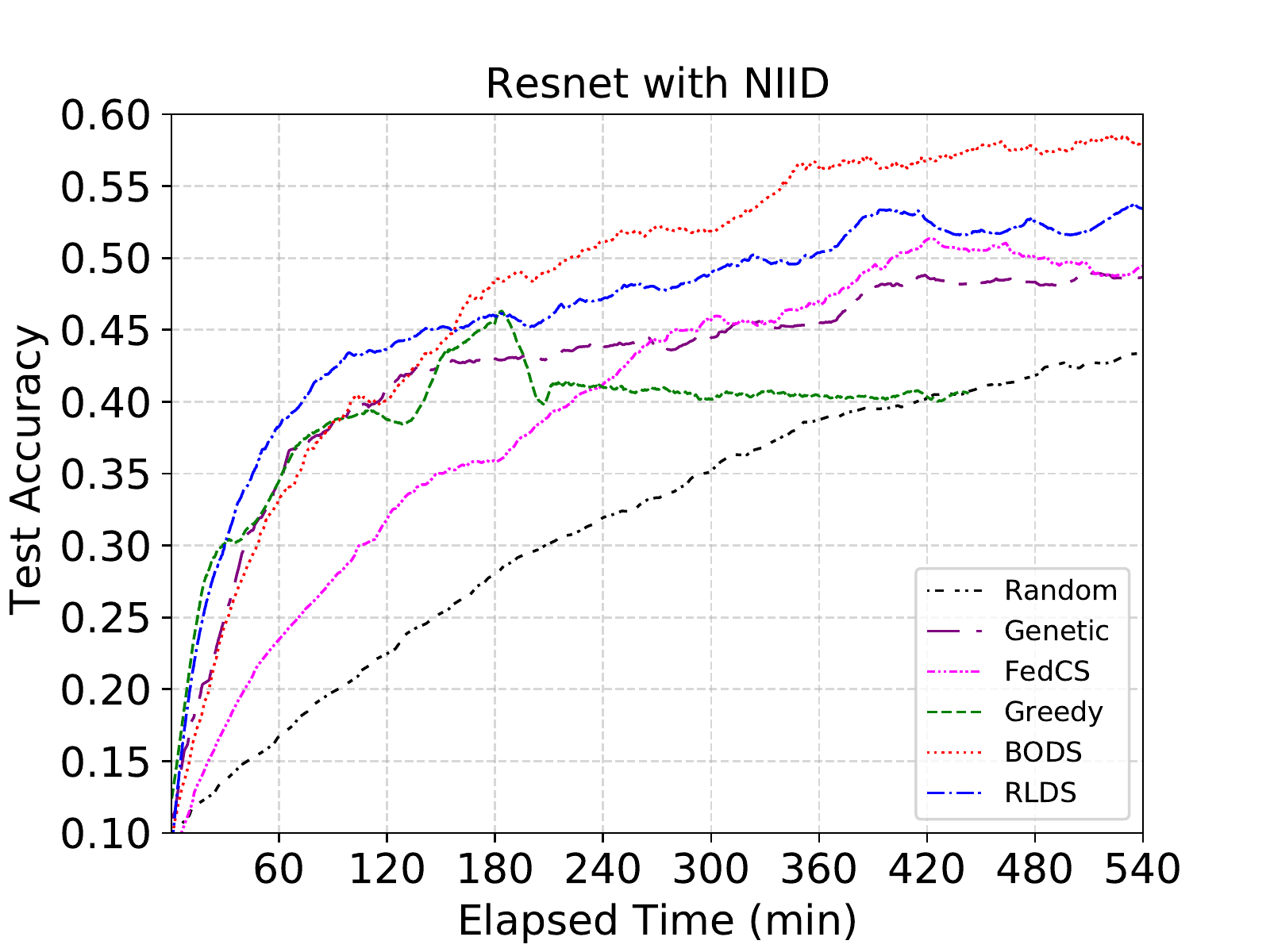}
\label{figReNiid}
\vspace{-4mm}
\caption{}
\end{subfigure}
\begin{subfigure}{0.3\linewidth}
\includegraphics[width=\linewidth]{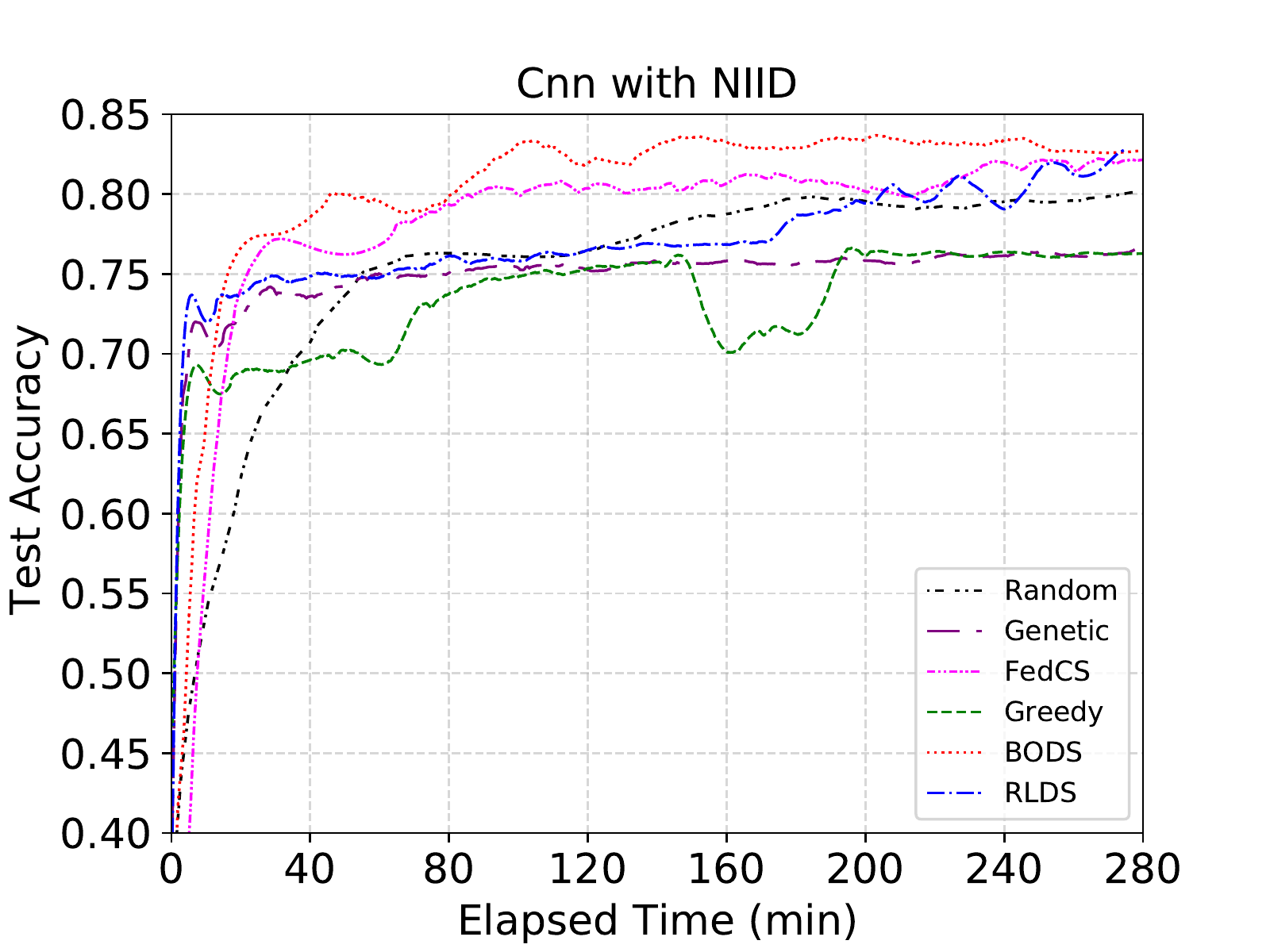}
\label{figCnNiid}
\vspace{-4mm}
\caption{}
\end{subfigure}
\begin{subfigure}{0.3\linewidth}
\includegraphics[width=\linewidth]{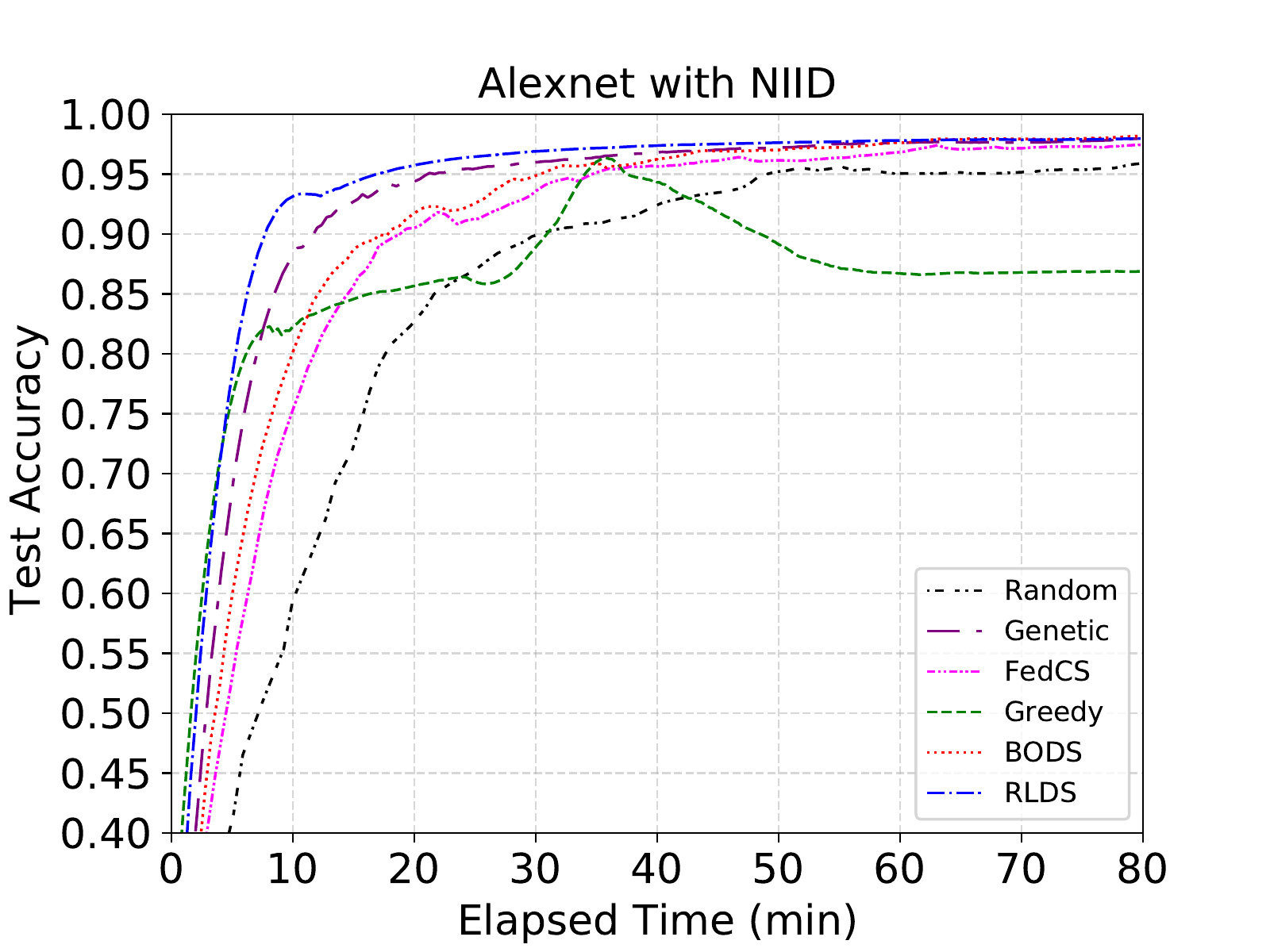}
\label{figAlNiid}
\vspace{-4mm}
\caption{}
\end{subfigure}
\begin{subfigure}{0.3\linewidth}
\includegraphics[width=\linewidth]{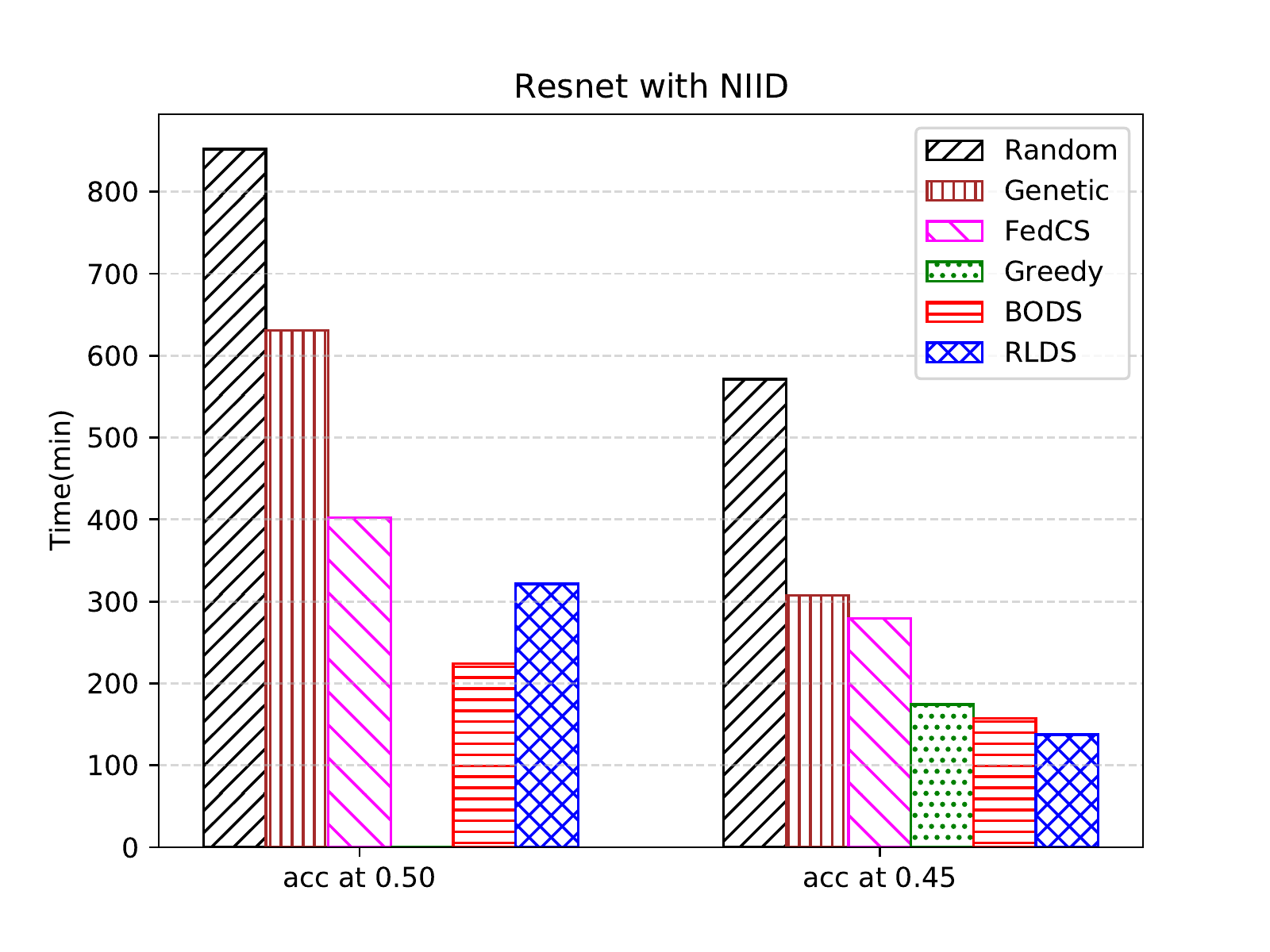}
\label{figReNiidAcc}
\vspace{-5mm}
\caption{}
\end{subfigure}
\begin{subfigure}{0.3\linewidth}
\includegraphics[width=\linewidth]{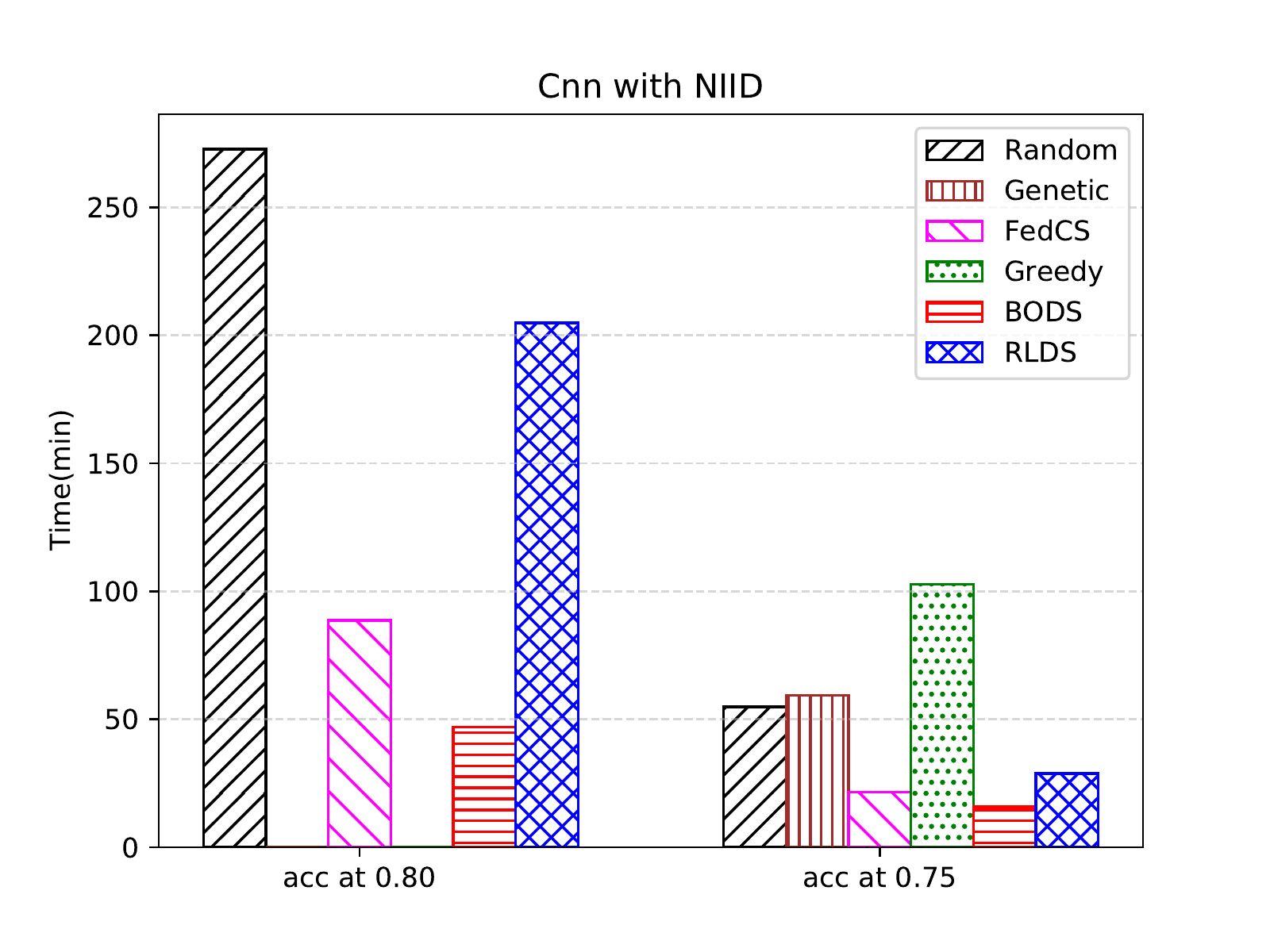}
\label{figCnNiidAcc}
\vspace{-5mm}
\caption{}
\end{subfigure}
\begin{subfigure}{0.3\linewidth}
\includegraphics[width=\linewidth]{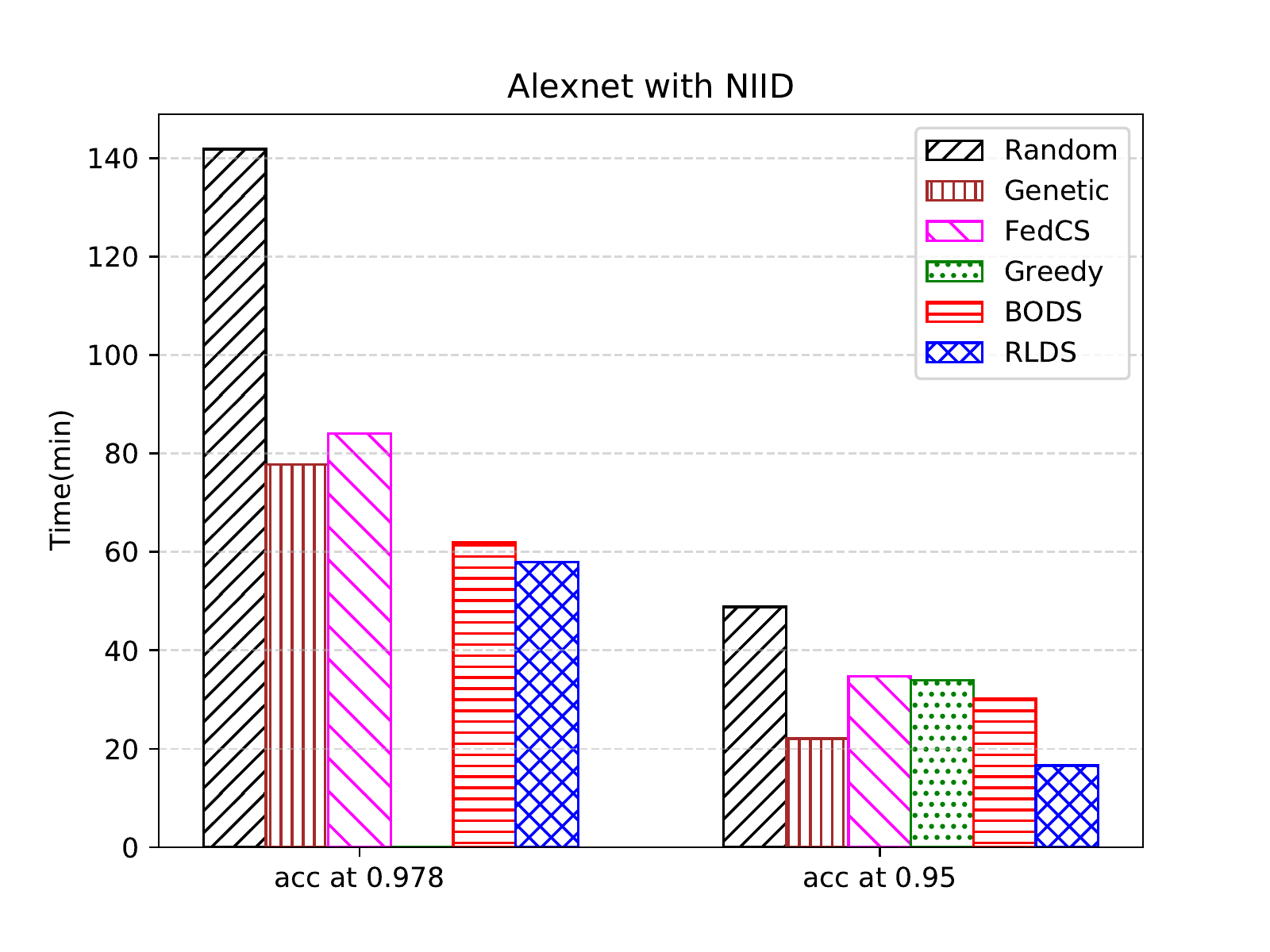}
\label{figAlNiidAcc}
\vspace{-5mm}
\caption{}
\end{subfigure}
\vspace{-2mm}
\caption{The convergence accuracy of different jobs in Group B changes over time with the non-IID distribution.  (d) to (f) show the time required for scheduling methods to achieve the target accuracy with non-IID, where Greedy and Genetic fail to achieved the target accuracy on some jobs.}
\label{fig:groupBNonIID}
\end{figure*}

\begin{figure*}[htbp]
\centering
\begin{subfigure}{0.3\linewidth}
\includegraphics[width=\linewidth]{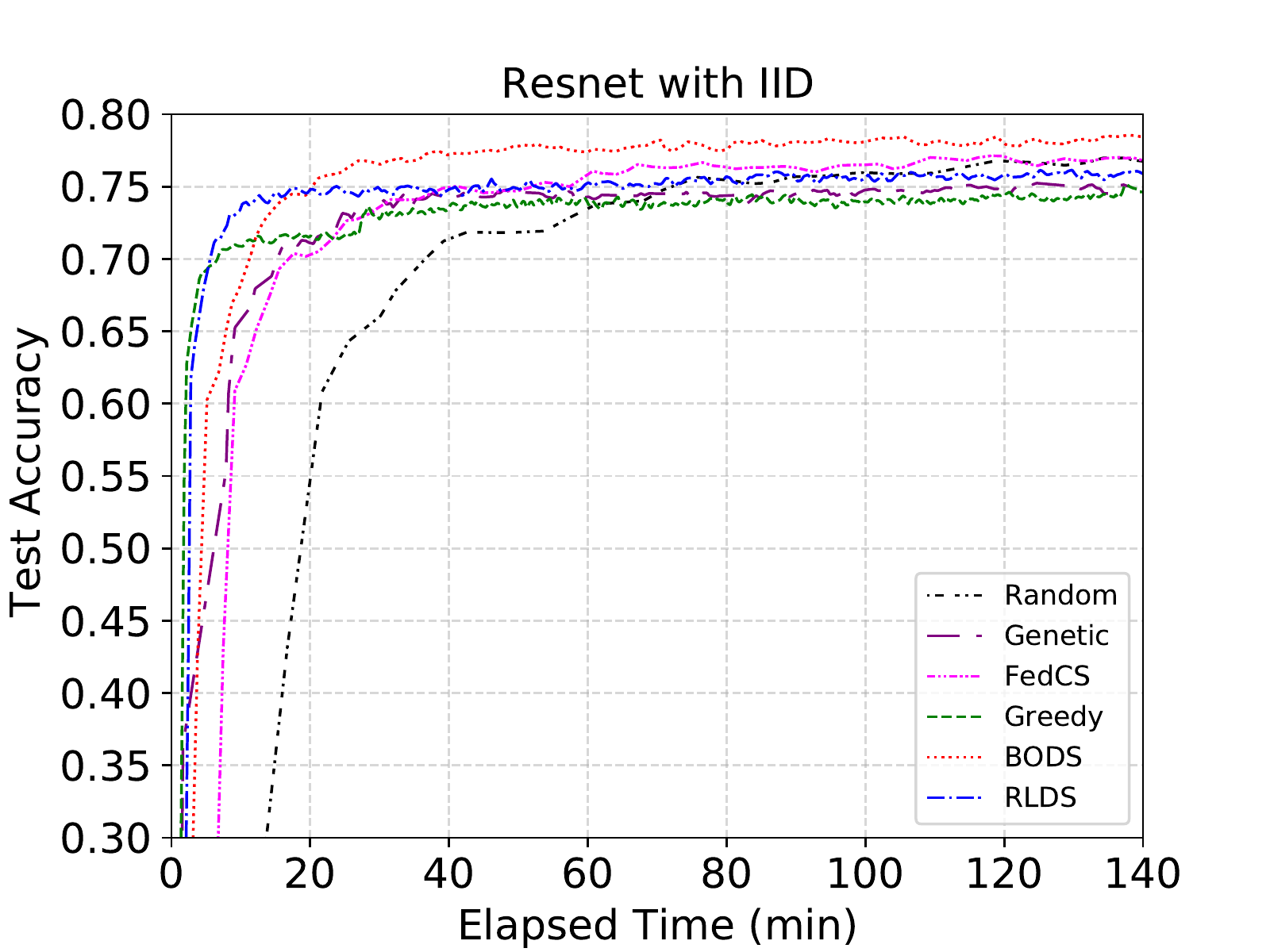}
\label{figReiid}
\vspace{-4mm}
\caption{}
\end{subfigure}
\begin{subfigure}{0.3\linewidth}
\includegraphics[width=\linewidth]{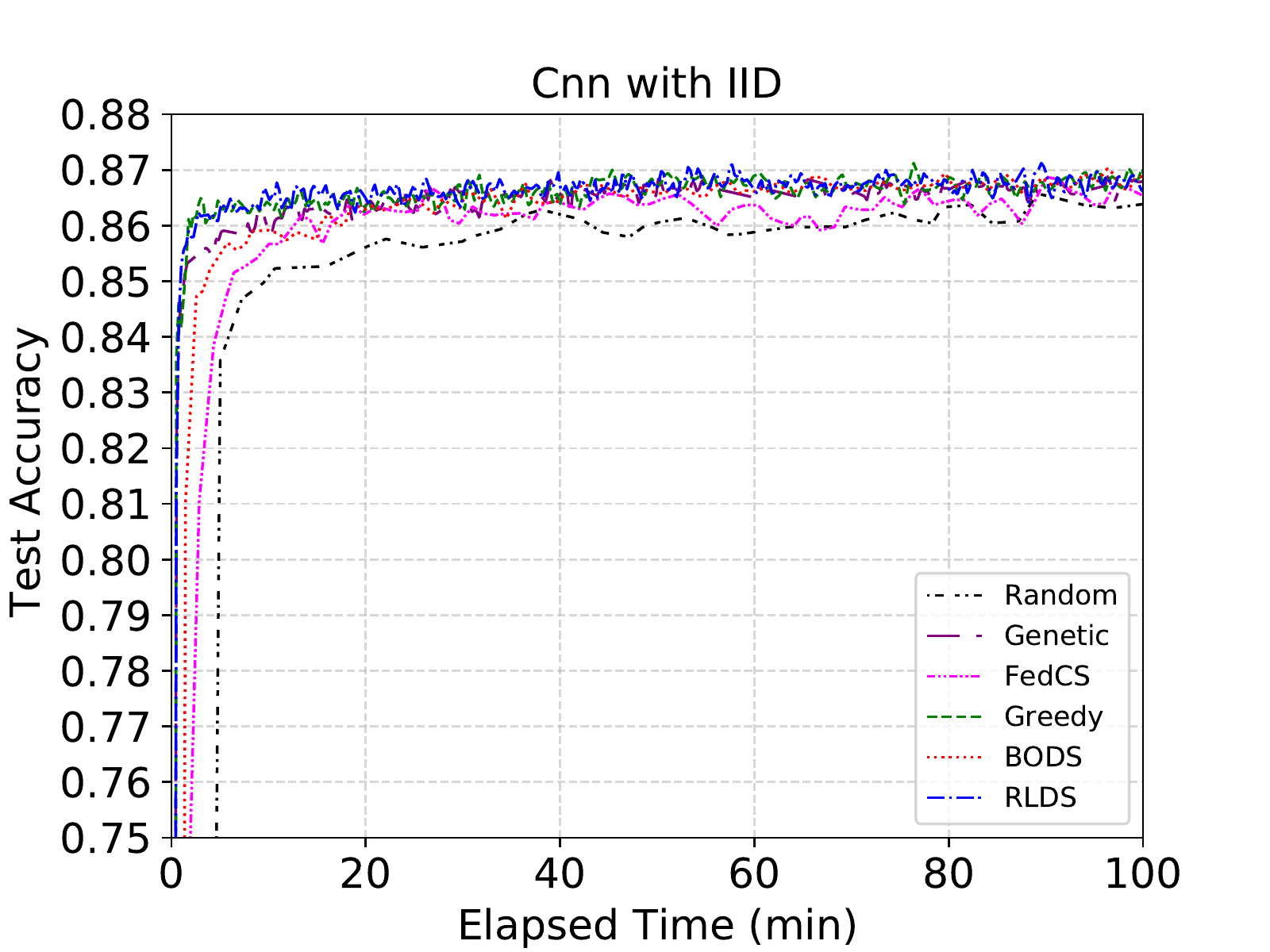}
\label{figCniid}
\vspace{-4mm}
\caption{}
\end{subfigure}
\begin{subfigure}{0.3\linewidth}
\includegraphics[width=\linewidth]{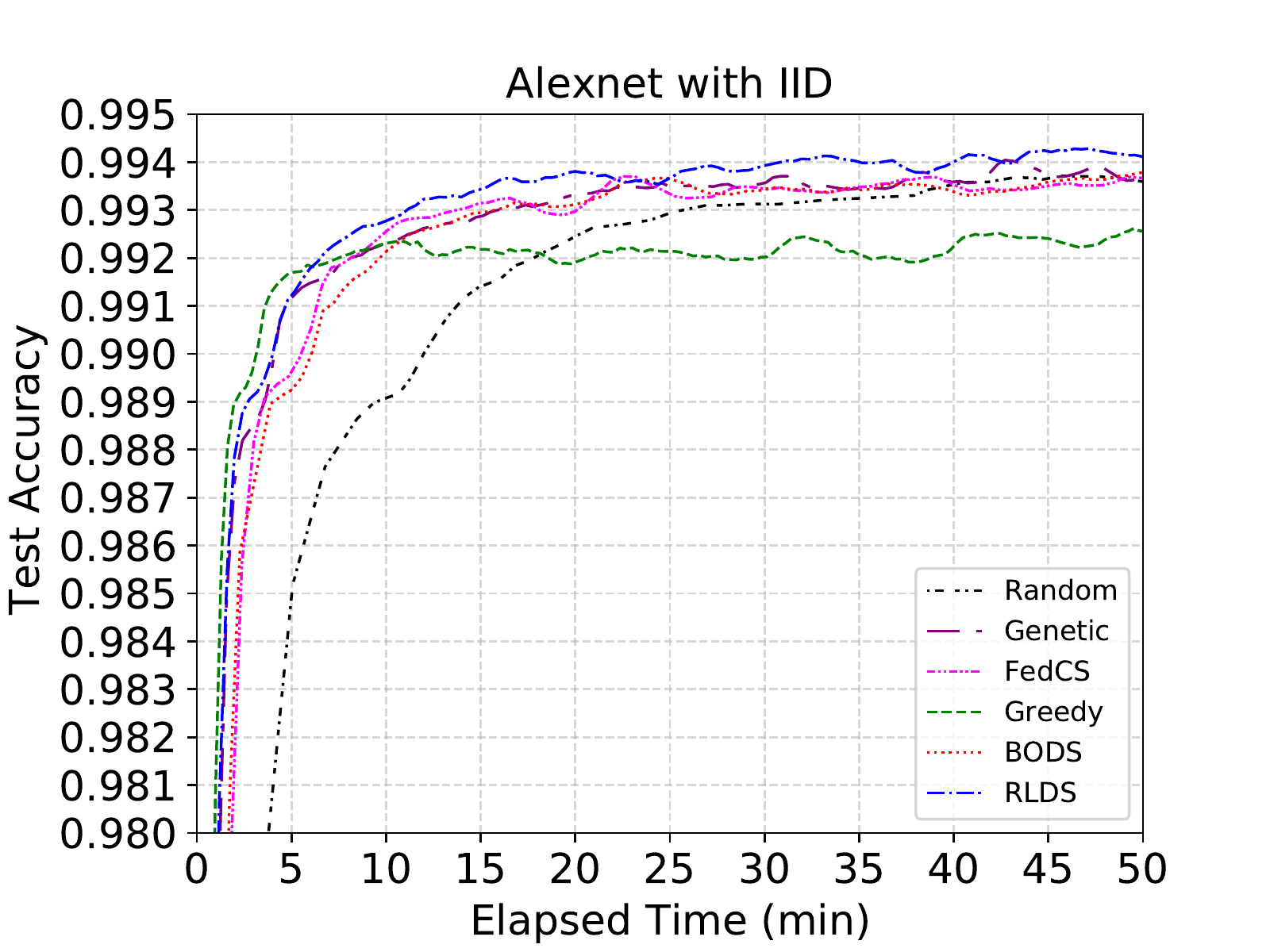}
\label{figAliid}
\vspace{-4mm}
\caption{}
\end{subfigure}
\begin{subfigure}{0.3\linewidth}
\includegraphics[width=\linewidth]{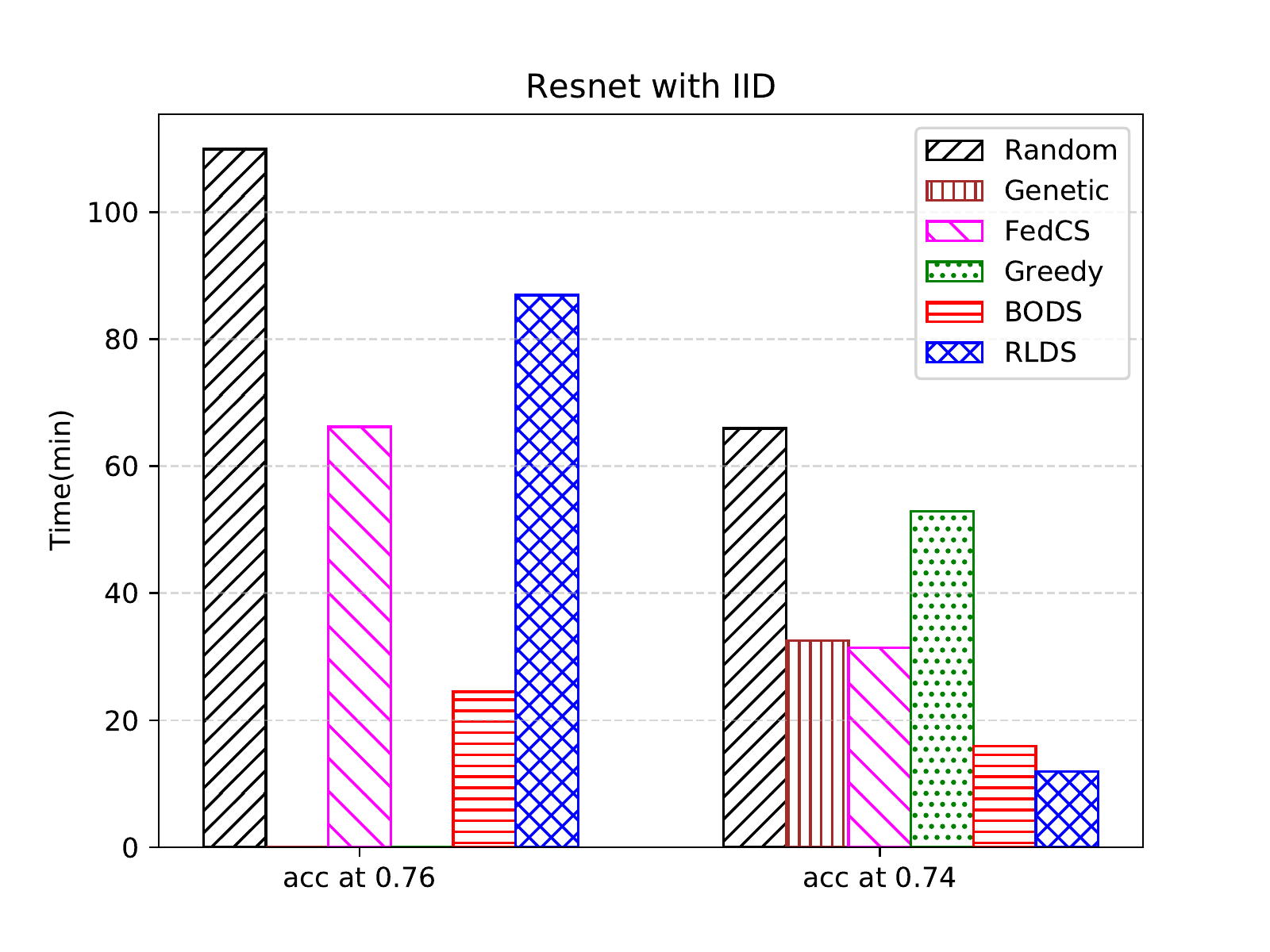}
\label{figReiidAcc}
\vspace{-5mm}
\caption{}
\end{subfigure}
\begin{subfigure}{0.3\linewidth}
\includegraphics[width=\linewidth]{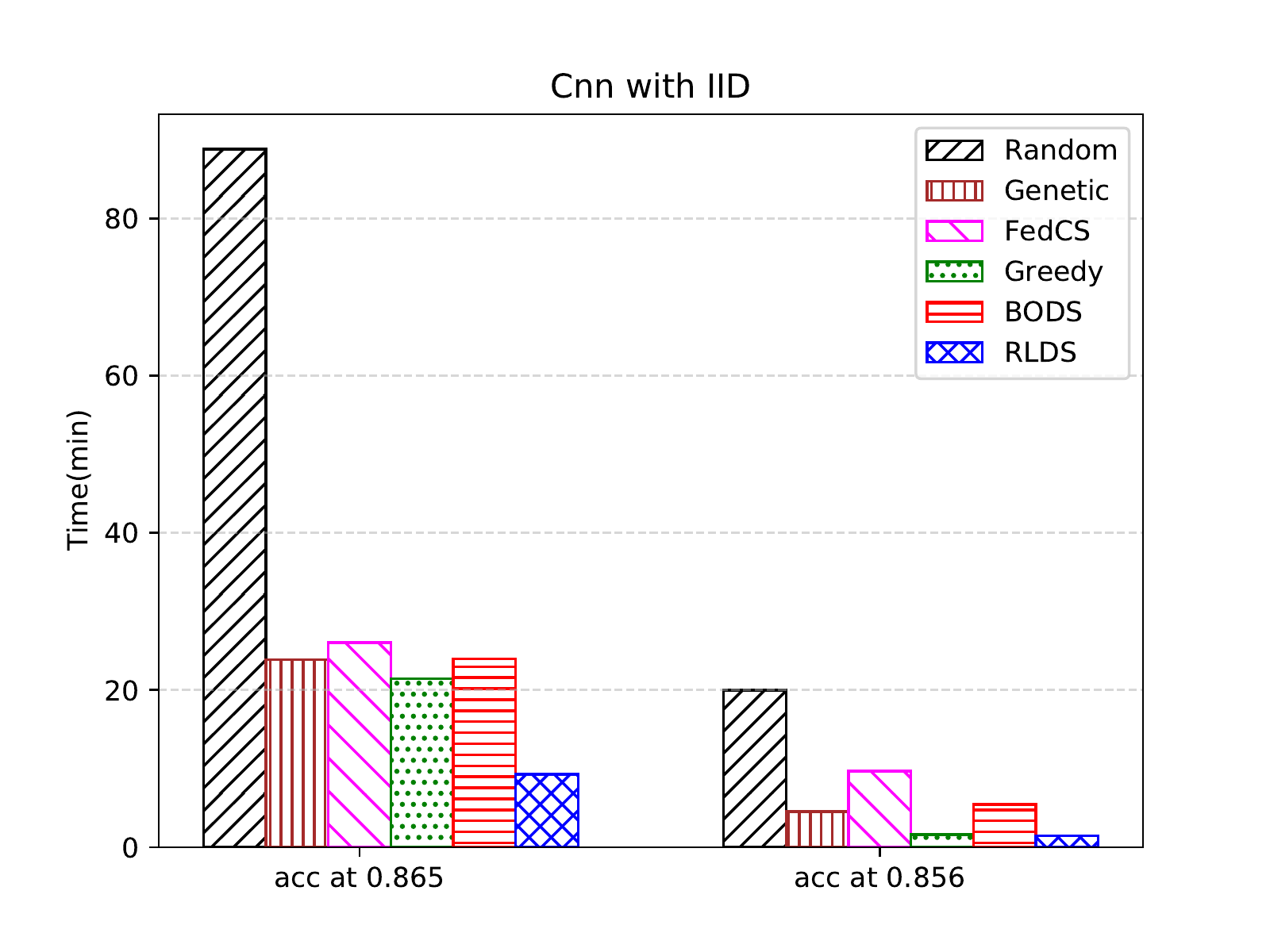}
\label{figCniidAcc}
\vspace{-5mm}
\caption{}
\end{subfigure}
\begin{subfigure}{0.3\linewidth}
\includegraphics[width=\linewidth]{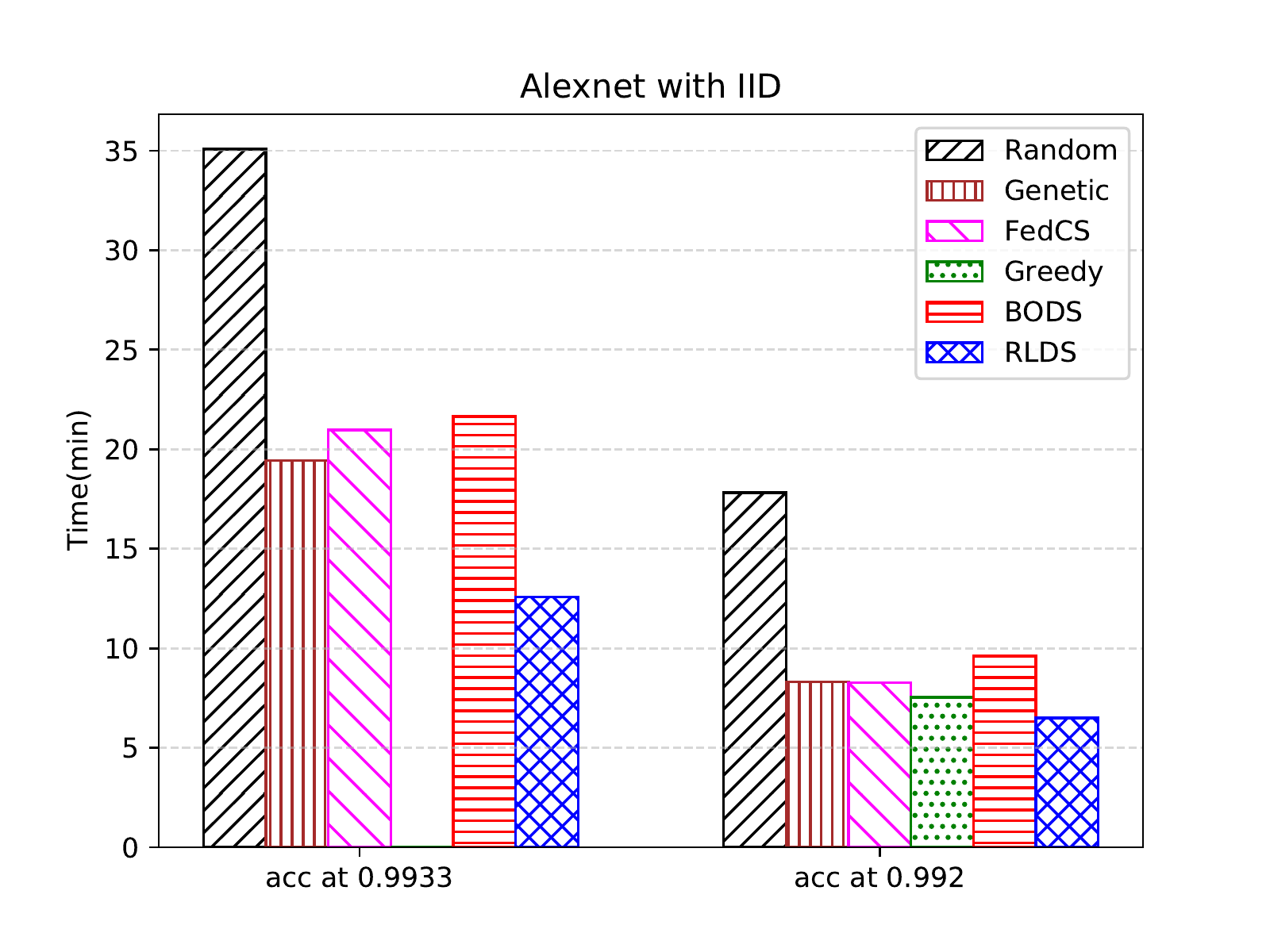}
\label{figAliidAcc}
\vspace{-5mm}
\caption{}
\end{subfigure}
\vspace{-2mm}
\caption{The convergence accuracy of different jobs in Group B changes over time with the IID distribution. (d) to (f) show the time required for scheduling methods to achieve the target accuracy with IID, where Greedy and Genetic fail to achieve the target accuracy on some jobs.}
\label{fig:detail-IID-GroupB}
\end{figure*}

Figures \ref{fig:groupBNonIID} and \ref{fig:detail-IID-GroupB} show the obvious fast convergence speed of RLDS and BODS. We conducted the experiment with different target accuracy, and the advantage is up to 7.63 times faster for Target 1 (0.845), 9.89 times faster for Target 2 (0.856), and 6.98 times faster for Target 3 (0.865), compared with the baselines. We conduct an ablation experiment and find that the data fairness improves both the convergence speed (up to 9.35 times faster) and the accuracy (up to 15.3\%). In addition, other scheduling methods, e.g., simulated annealing \cite{van1987simulated}, corresponds to worse performance (up to 91.4\% slower and 3.5\% lower accuracy), according to our experiments. We carry out experiments to compare other black-box optimization methods (deep neural networks \cite{zang2019hybrid}), of which the performance is worse (up to 90.5\% slower and 26.3\% lower accuracy) than our methods. Furthermore, we tested other combinations of the two costs, which correspond to worse performance (up to 37.1\% slower and 3.5\% lower accuracy for the sum of squared costs, and 64.4\% slower and 3.3\% lower accuracy for multiplication) compared to the linear one (Formula \ref{eq:totalCost}). 

\subsection{Design Details}

As an FL environment may contain GPUs or other high-performance chips, it is beneficial to train multiple jobs simultaneously to reduce training time while achieving the same accuracy. Within each round, Step \textcircled{6} exploits FedAvg \cite{McMahan2017Communication-efficien} to aggregate multiple models within each job, which can ensure the optimal convergence \cite{li2019convergence, Zhou2018On}. Within our framework, the sensitive raw data is kept within each device, while only the models are allowed to be transferred. Other methods, e.g., homomorphic encryption \cite{paillier1999public} and differential privacy \cite{dwork2008differential}, can be exploited to protect the privacy of sensitive data. 

We choose the linear combination because of its convenience and good performance. In practice, we empirically set $\alpha$ and $\beta$ based on the information from previous execution and adjust them using small epochs. We increase $\alpha$ for fast convergence and increase $\beta$ mainly for high accuracy.

Please note that the ``data fairness'' is different from the ``fairness'' (the bias of the machine learning models concerning certain features) in machine learning \cite{mehrabi2021survey}. Formula \ref{eq:fairness} is based on \cite{petrangeli2014multi}, and  we are among the first to extend this idea from distributed or network systems to FL. When the devices are non-uniformly sampled with low data fairness, the convergence is slowed down \cite{li2019convergence, Zhou2018On}. In addition, data fairness is important due to the underlying data heterogeneity across the devices. Data fairness can help arbitrarily select devices without harming the learning performance.

The BO-based and RL-based methods are designed for different model complexities, and we choose the better one based on known profiling information with small tests (a few epochs) to avoid possible limitations. RLDS favors complex jobs, as it can learn the influence among diverse devices. The influence refers to the concurrent, complementary, and latent impacts of the data in multiple devices for diverse jobs. However, BODS favors simple jobs, while it relies on simple statistical knowledge. The complexity of jobs is determined by the number of parameters of models and the size of the training dataset.

In fact, we consider the probability to release the devices in $\mathcal{V}_{o}$ in BODS and RLDS, and possible concurrent occupation of other devices for other jobs, which is not explained in the paper to simplify the explanation.

During the execution, we only sample 10\% devices of all the devices for each job. Thus, we do not assume that all the devices are available all the time during the training process.

\begin{figure*}[htbp]
\centering
\begin{subfigure}{0.3\linewidth}
\includegraphics[width=\linewidth]{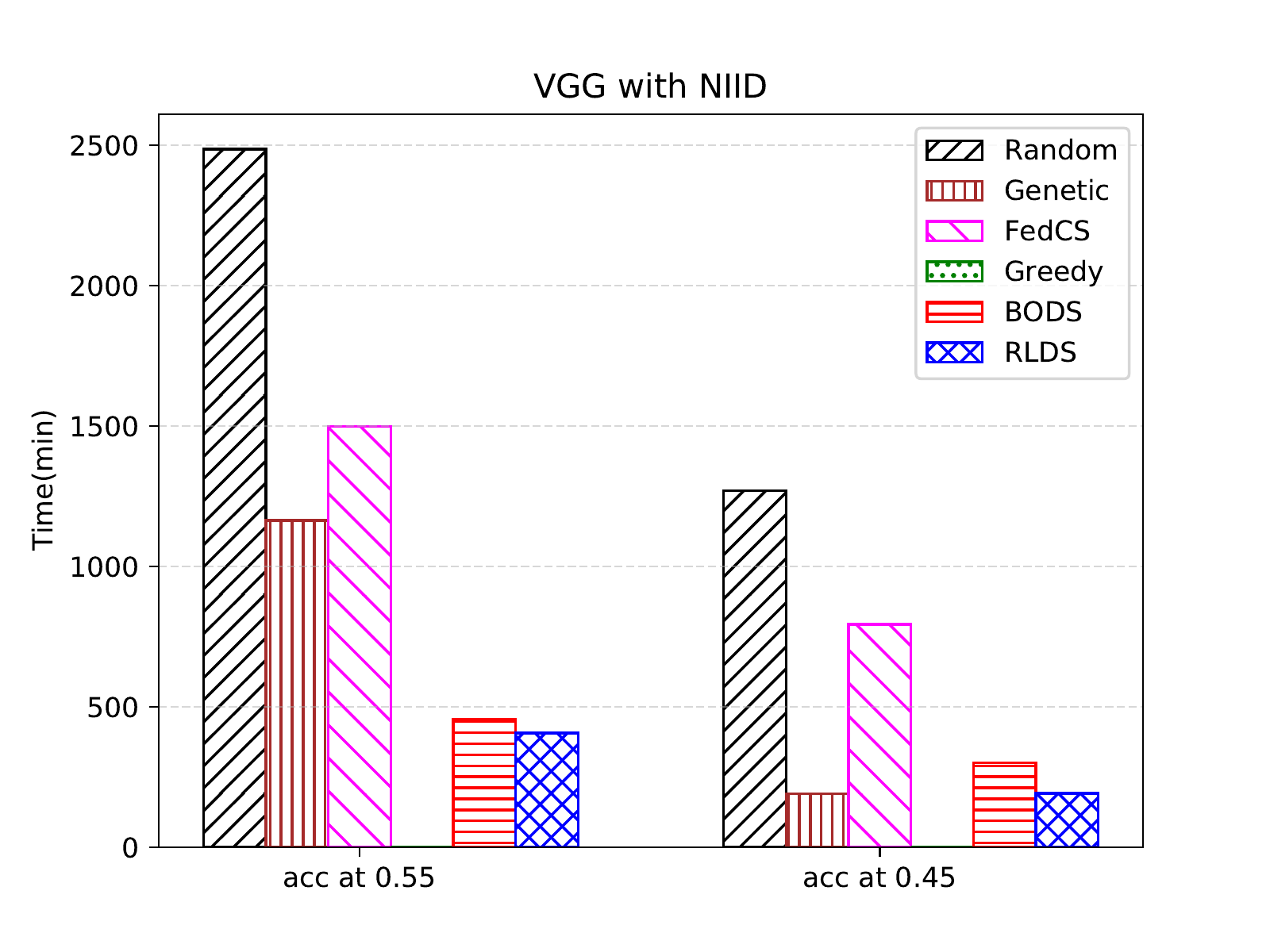}
\label{figVgNiidAcc}
\vspace{-5mm}
\caption{}
\end{subfigure}
\begin{subfigure}{0.3\linewidth}
\includegraphics[width=\linewidth]{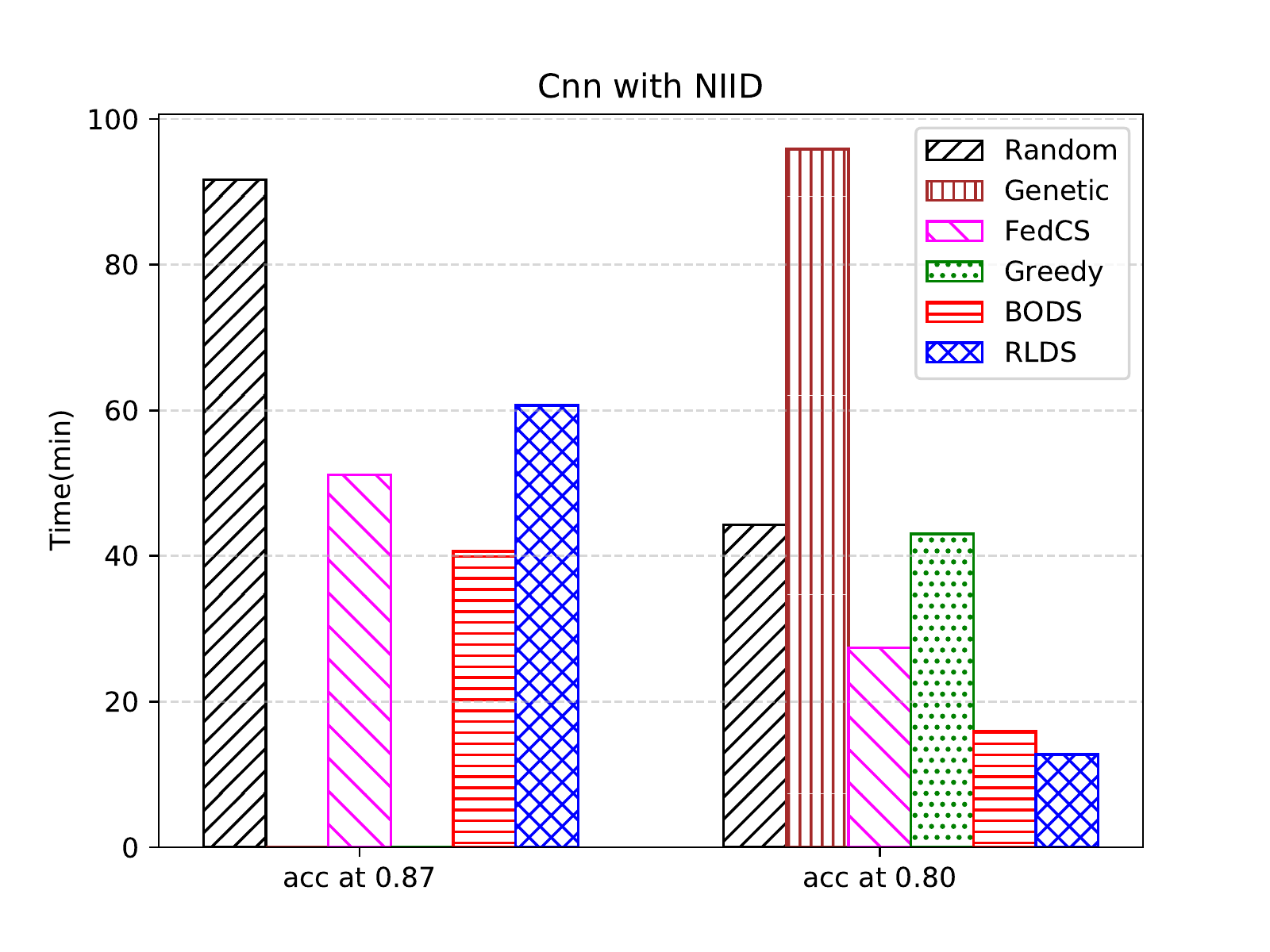}
\label{figCNiidAcc}
\vspace{-5mm}
\caption{}
\end{subfigure}
\begin{subfigure}{0.3\linewidth}
\includegraphics[width=\linewidth]{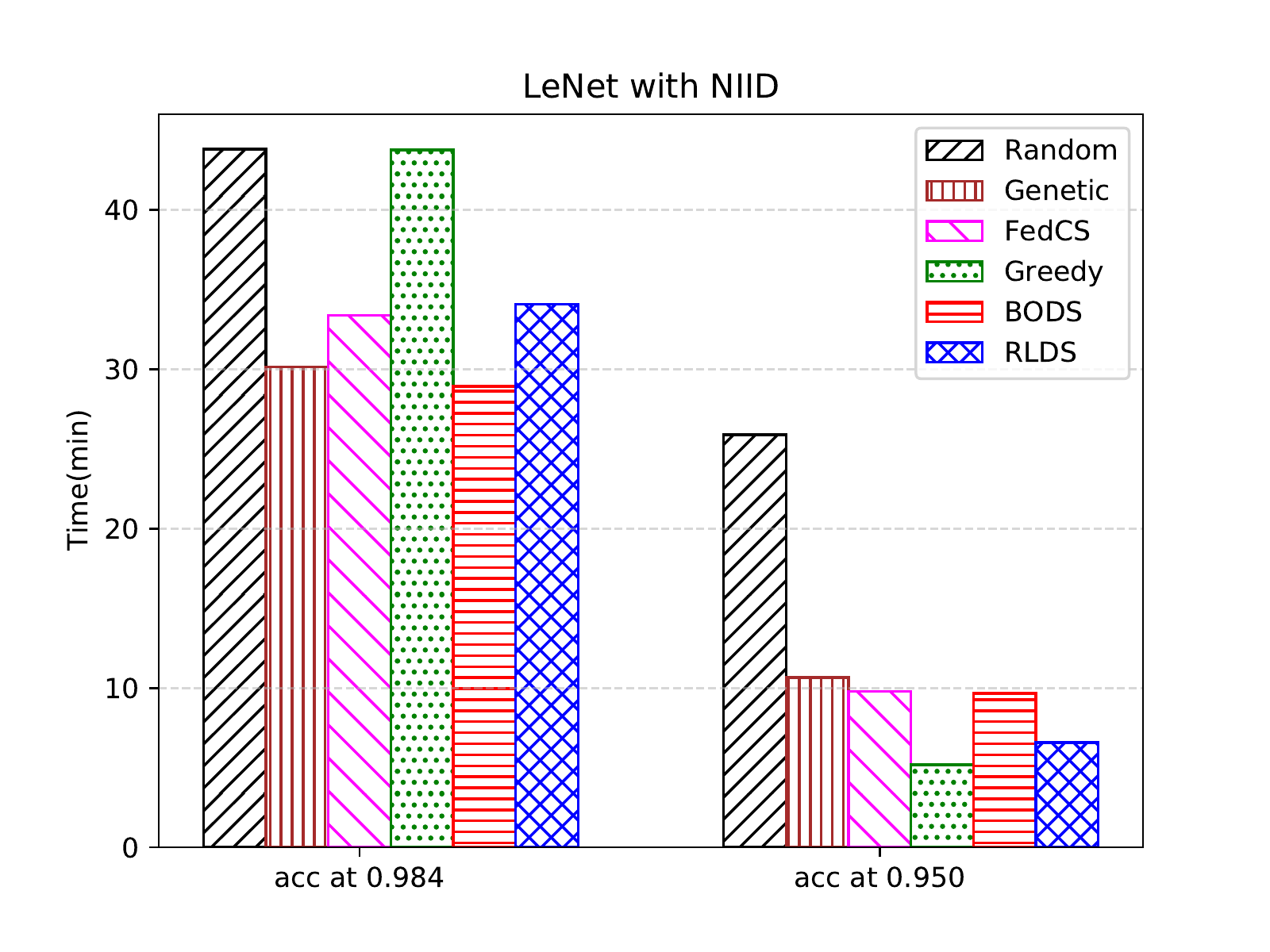}
\label{figLeNiidAcc}
\vspace{-5mm}
\caption{}
\end{subfigure}
\vspace{-2mm}
\caption{The time required for each job of Group A to achieve the target convergence accuracy with the non-IID distribution. As Greedy and Genetic fail to achieve the target accuracy on some jobs, the time is not shown.}
\label{fig:time-non-IID-Group-A}
\end{figure*}

\begin{figure*}[htbp]
\centering
\begin{subfigure}{0.3\linewidth}
\includegraphics[width=\linewidth]{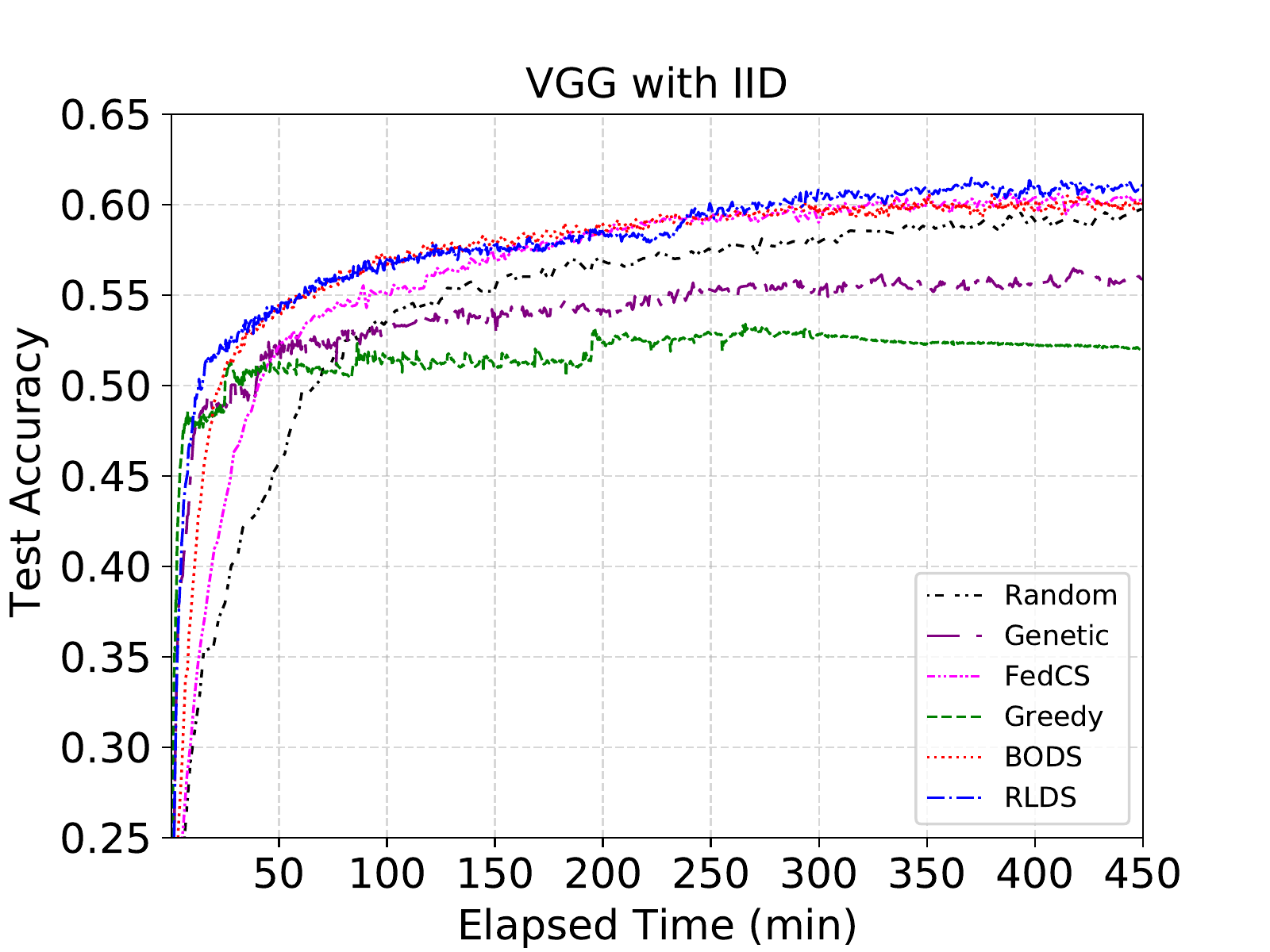}
\label{figVgiid}
\vspace{-4mm}
\caption{}
\end{subfigure}
\begin{subfigure}{0.3\linewidth}
\includegraphics[width=\linewidth]{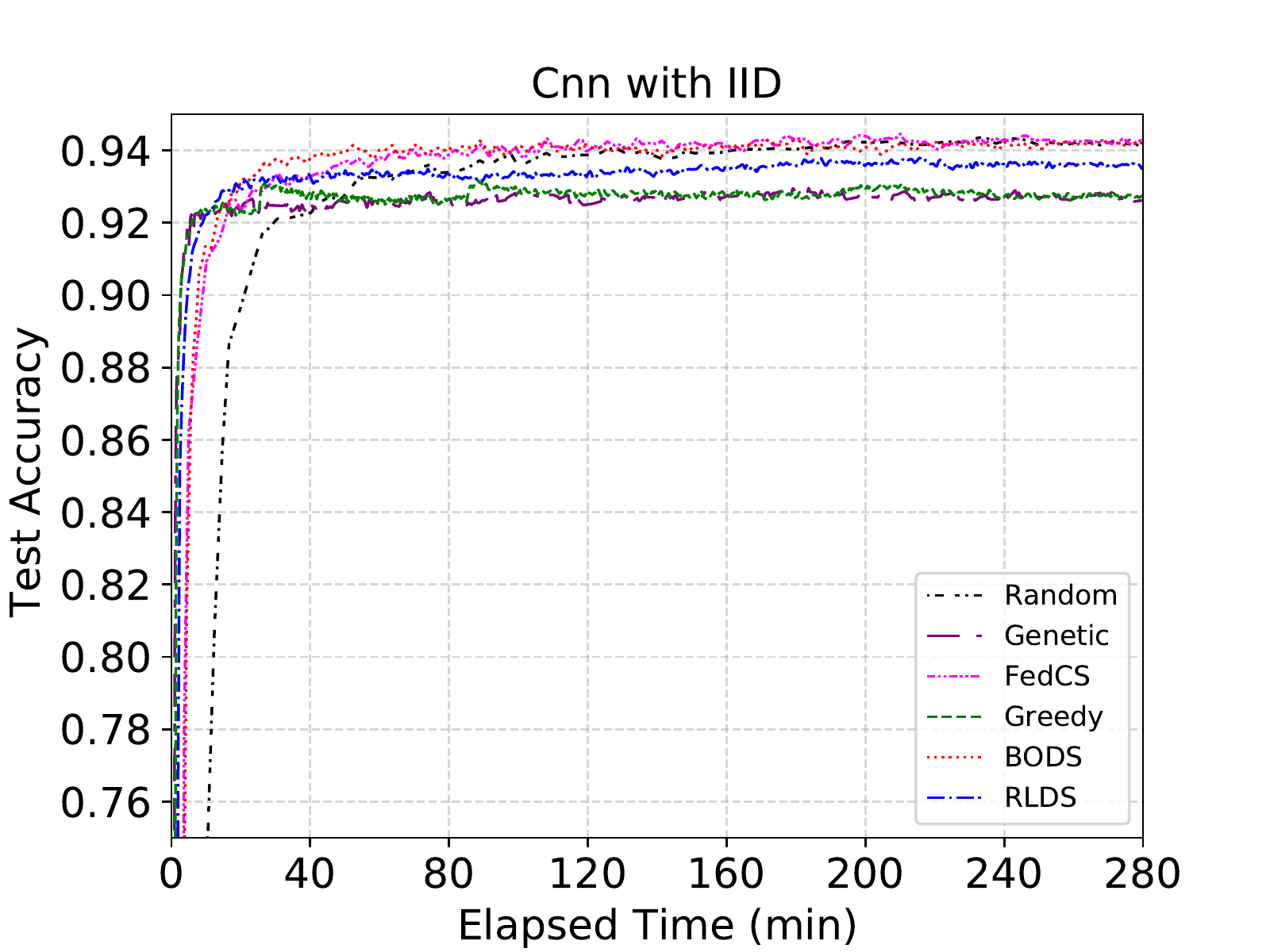}
\label{figCiid}
\vspace{-4mm}
\caption{}
\end{subfigure}
\begin{subfigure}{0.3\linewidth}
\includegraphics[width=\linewidth]{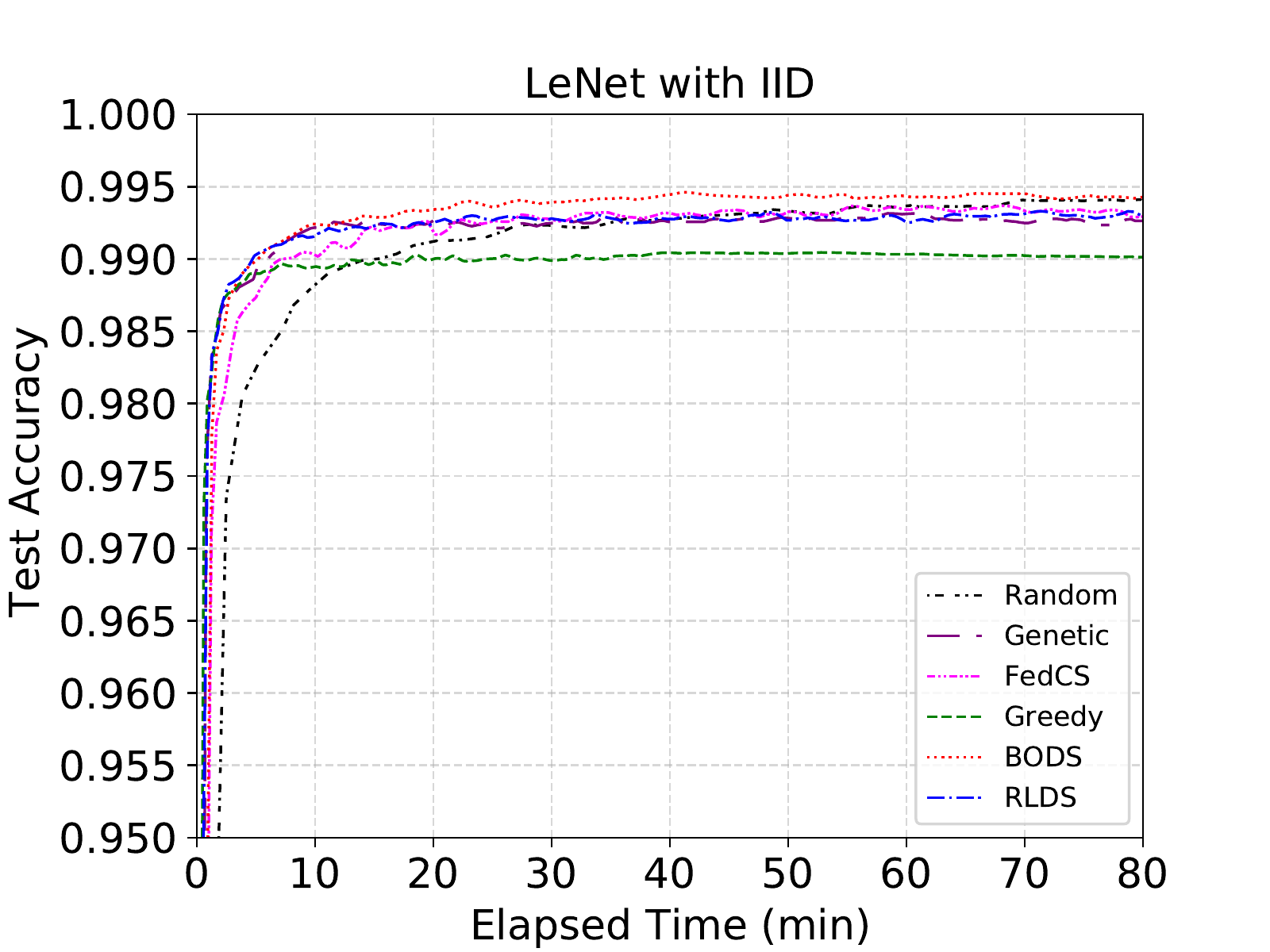}
\label{figLeiid}
\vspace{-4mm}
\caption{}
\end{subfigure}
\begin{subfigure}{0.3\linewidth}
\includegraphics[width=\linewidth]{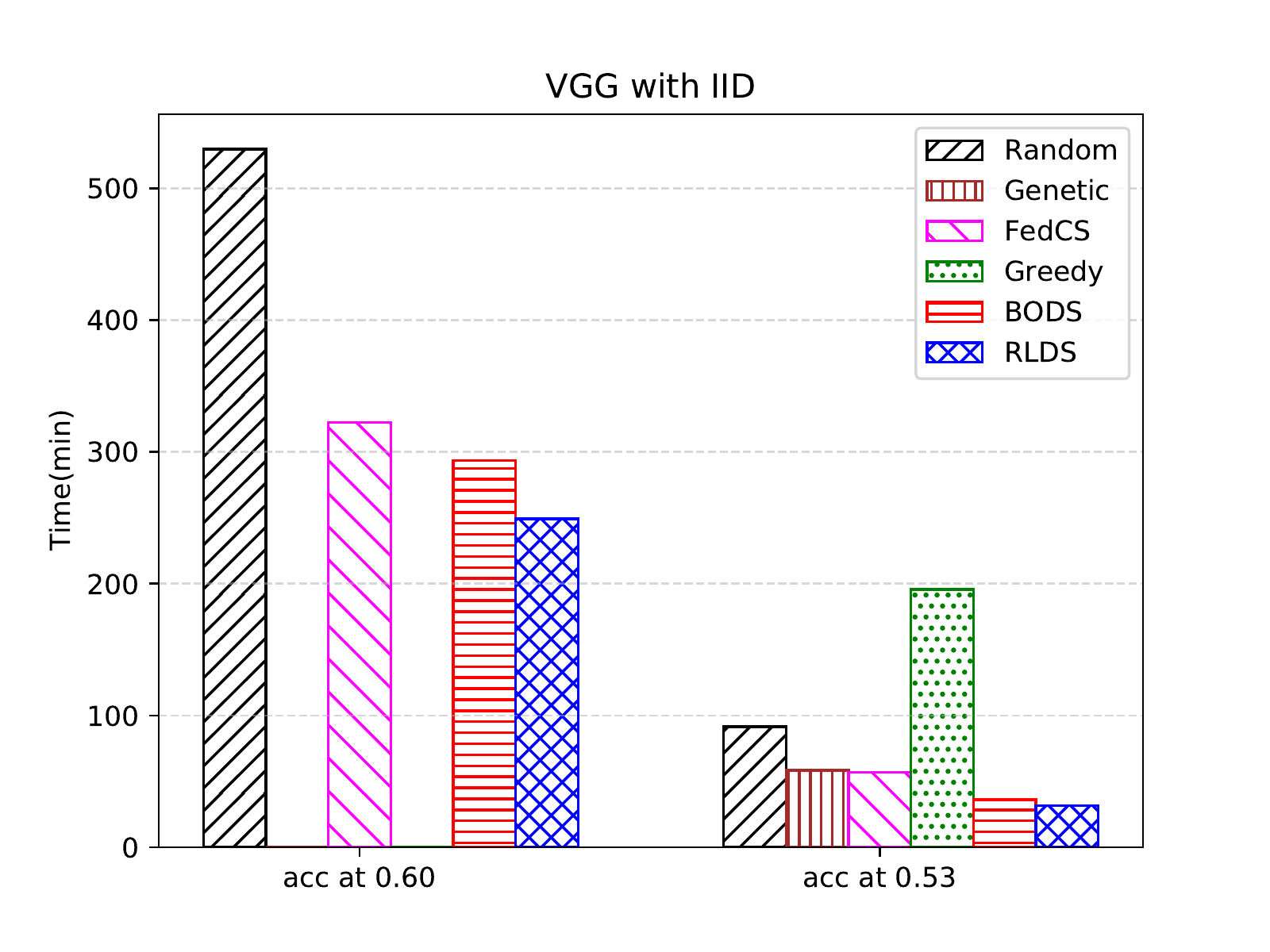}
\label{figVgiidAcc}
\vspace{-5mm}
\caption{}
\end{subfigure}
\begin{subfigure}{0.3\linewidth}
\includegraphics[width=\linewidth]{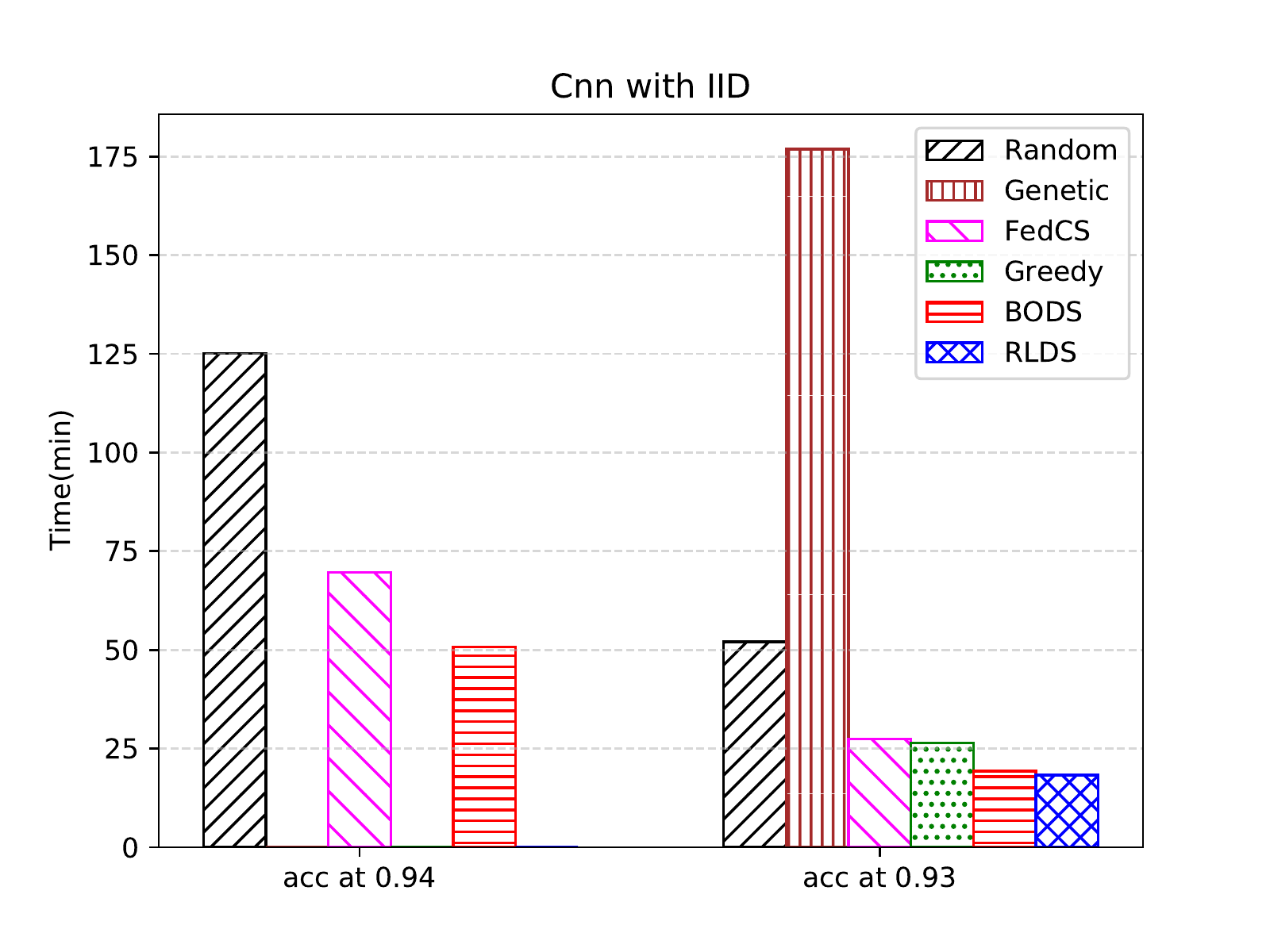}
\label{figCiidAcc}
\vspace{-5mm}
\caption{}
\end{subfigure}
\begin{subfigure}{0.3\linewidth}
\includegraphics[width=\linewidth]{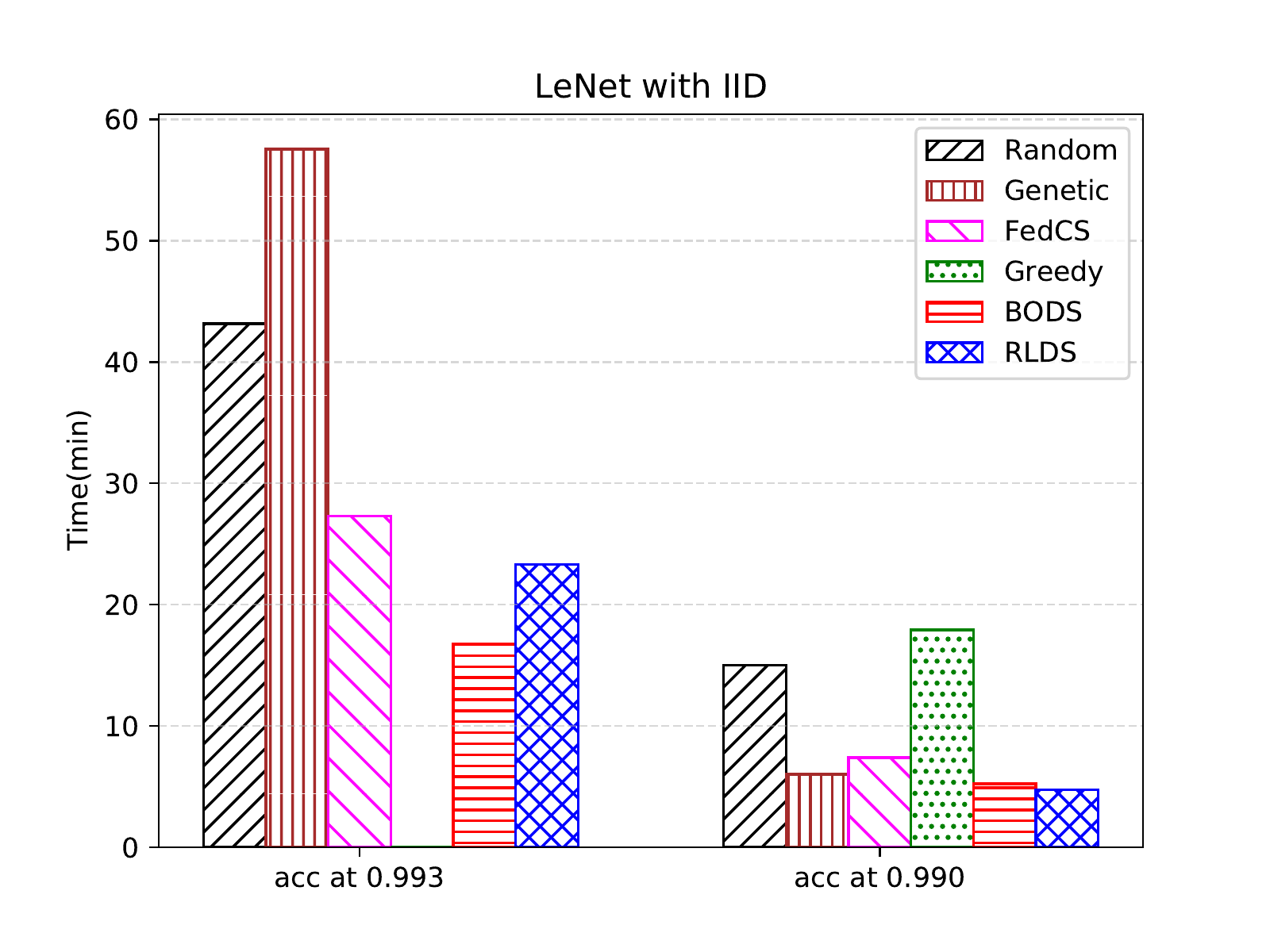}
\label{figLeiidAcc}
\vspace{-5mm}
\caption{}
\end{subfigure}
\vspace{-2mm}
\caption{The convergence accuracy of different jobs in Group A changes over time with the IID distribution.  (d) to (f) show the time required for scheduling methods to achieve the target accuracy with non-IID, where Greedy and Genetic fail to achieve the target accuracy on some jobs.}
\label{fig:detail-non-IID-GroupA}
\end{figure*}

\end{document}